\documentclass[twocolumn]{raa} 
\usepackage{graphicx,times}             %for PS/EPS graphics inclusion, new
\usepackage{natbib}
\usepackage{amssymb,amsmath}
\usepackage{multirow}
\usepackage{threeparttable}
\bibpunct{(}{)}{;}{a}{}{,}

\usepackage[pagebackref=true]{hyperref}

\begin{document}
	
	\title{Closeby Habitable Exoplanet Survey (CHES). V. Planetary Parameters Derived from Angular Separation Variations
	}

	\volnopage{Vol.0 (20xx) No.0, 000--000}      %%preserved for Editor. DOn't remove!
	\setcounter{page}{1}          %%starting page, preserved for Editor. DOn't remove!
		
	\author{ Dongjie Tan$^{1,2}$
		\and Jianghui Ji$^{1,2,3}$
		\and Chunhui Bao$^{1,2}$
		\and Xiumin Huang$^{1,2}$
		\and Guo Chen$^{1,2}$
		\and Su Wang$^{1,3}$
		\and Yao Dong$^{1,3}$
		\and Jiacheng Liu$^{4}$
		\and Zi Zhu$^{4,5}$
		\and Haitao Li$^{6,7,8}$
		\and Junbo Zhang$^{8,9,10}$
		\and Liang Fang$^{8,10}$
		\and Dong Li$^{11}$
		\and Lei Deng$^{11}$
	}
	
	\institute{CAS Key Laboratory of Planetary Sciences, Purple Mountain Observatory, Chinese Academy of Sciences,
		Nanjing 210023, China; {\it jijh@pmo.ac.cn}\\
		\and
		School of Astronomy and Space Science, University of Science and Technology of China, Hefei 230026, China\\
		\and
		CAS Center for Excellence in Comparative Planetology, Hefei 230026, China\\
		\and
		School of Astronomy and Space Science, Nanjing University, Nanjing 210046, China\\
		\and
		University of Chinese Academy of Sciences, Nanjing 211135, China\\
		\and
		National Space Science Center, Chinese Academy of Sciences, Beijing 100190, China\\
		\and
		Department of Physics and Astronomy, Faculty of Environment Science and Economy, University of Exeter, EX4 4QL, UK\\
		\and
		University of Chinese Academy of Sciences, Beijing 100049, China\\
		\and
		China National Laboratory on Adaptive Optics, Chengdu 610209, China\\
		\and
		Institute of Optics and Electronics, Chinese Academy of Sciences, Chengdu 610209, China\\
		\and
		Innovation Academy for Microsatellites of Chinese Academy of Sciences, Shanghai 201306, China\\
		\vs\no
		{\small Received 20xx month day; accepted 20xx month day}}

	\abstract{The Closeby Habitable Exoplanet Survey (CHES) aims to achieve microarcsecond-level astrometry of about one hundred nearby FGK-type stars within 10 parsecs to detect Earth-like planets. Such precision exceeds the capability of absolute astrometry relying on \emph{Gaia} catalogs, whose positional accuracy degrades over time due to error propagation from stellar motion and epoch offsets, limiting their use in microarcsecond-level detection. Traditional relative astrometry depends on positional components along right ascension and declination, requiring precise knowledge of field rotation and satellite attitude, which introduces additional errors. To address this, we propose a new relative measurement model based solely on variations in the length of angular separation between the target and reference stars, independent of direction. The model incorporates effects such as proper motion, parallax, radial velocity, light aberration, gravitational lensing, and planetary perturbations, enabling reconstruction of planetary orbits and masses. This approach enhances measurement stability and precision, providing a framework that is not entirely dependent on the \emph{Gaia} catalog and suitable for CHES and other future high-accuracy astrometric missions.
	}
	
	\authorrunning{Tan et al.}            %author_head in even pages
	\titlerunning{Planetary Parameters Derived from Angular Separation Variations}  % title_head in odd pages
	
	\maketitle

	\section{Introduction}
	\label{sec:intro}
	
	Astrometry, the branch of astronomy concerned with measuring celestial positions and motions, has been fundamental since antiquity. Its development has closely followed advances in instrumentation, from naked-eye observations to space-based high-precision measurements, significantly improving our understanding of celestial dynamics and cosmic evolution \citep{Perryman2012}. With missions such as Hipparcos \citep{vanLeeuwen2007b} and \emph{Gaia} \citep{GaiaCollaboration2016}, astrometry has entered a new era \citep{Brown2021}. In particular, \emph{Gaia} has provided unprecedented measurements of stellar positions and motions, enabling the detection of binary systems and black holes, including several candidates identified through astrometric signals \citep{El-Badry2023a, El-Badry2023b, GaiaCollaboration2024}. Its Data Release 3 achieves a precision of $\sim20$ microarcseconds \citep{Lindegren2021}.
	
	These advances have opened a new avenue for exoplanet detection through astrometry. Traditional methods, such as radial velocity \citep{Vogt1994} and transit observations \citep{Howell2014, Ricker2015}, have achieved significant success but remain subject to limitations related to orbital configuration and observational constraints. Astrometry, by measuring stellar reflex motion, directly probes gravitational perturbations induced by orbiting bodies. With microarcsecond-level precision \citep{Malbet2012}, it provides a promising approach for detecting Earth-like planets. To date, three exoplanets have been identified astrometrically \citep{Sahlmann2013, Curiel2022}, including one from \emph{Gaia} observations \citep{Sozzetti2023}.
	
	In addition to \emph{Gaia}, several space missions aim to advance astrometry and exoplanet studies through complementary approaches. The JASMINE series focuses on infrared astrometry of the Galactic center, enabling observations in high-extinction regions \citep{Gouda2012, Kawata2023}, with missions such as Nano-JASMINE and Small-JASMINE \citep{Yamada2013, Gouda2020}. High-precision differential astrometry missions, including NEAT and Theia \citep{Malbet2012,TheTheiaCollaboration2017}, aim to achieve sub-microarcsecond precision for detecting Earth-like planets. Other missions employ different techniques: PLATO \citep{Rauer2014} and ARIEL \citep{Tinetti2016} focus on transits and atmospheric characterization, while the Nancy Grace Roman Space Telescope \citep{Wang2022} and Euclid \citep{EuclidCollaboration2025} utilize microlensing. Direct imaging missions such as HabEx \citep{Gaudi2020} and the Habitable Worlds Observatory \citep{Mamajek2024} target direct detection of Earth-like planets. Despite these efforts, achieving microarcsecond precision for Earth-like planet detection remains challenging.
	
	Detecting Earth-like planets in habitable zones requires microarcsecond-level precision, beyond the current limits of the \emph{Gaia} catalog. Although future releases are expected to improve precision \citep{GaiaCollaboration2023}, the temporal propagation of proper motion and parallax uncertainties leads to increasing positional errors over time \citep{Gai2022}, limiting the capability of \emph{Gaia} for such detections.
	
	Therefore, traditional absolute astrometry based on prior catalogs becomes insufficient, and relative astrometric techniques are required. These methods measure positional variations of a target star relative to nearby reference stars. However, achieving high precision still requires accurate determination of displacements along right ascension and declination, or equivalently, along a fixed reference direction.
	
	In the CHES mission \citep{Ji2022, Ji2024, Bao2024a,Bao2025,Huang2025,Jiang2025}, a relative measurement strategy is adopted \citep{Tan2024}. Observations are performed over multiple epochs, during which field rotation due to spacecraft attitude changes is unavoidable. Unlike \emph{Gaia}, which derives global astrometric parameters through AGIS \citep{Lindegren2012,OMullane2011,Bombrun2012}, CHES focuses on targeted observations of nearby stars and cannot rely on a global reference frame. As a result, determining field rotation or maintaining microarcsecond precision in RA and Dec becomes a major challenge.
	
	To address this issue, we propose a new relative astrometric model based solely on variations in angular separation, without requiring knowledge of the orientation of the connecting line. This approach removes the dependence on field rotation and reference direction. With sufficient measurement precision, angular separations at the microarcsecond level can be achieved. The model includes proper motion, parallax, radial velocity, stellar aberration, gravitational lensing, and planetary perturbations. By fitting the observed variations, the stellar motion parameters can be recovered, and with multiple reference stars, the two-dimensional planetary signal can be reconstructed.
	
	This method is applicable not only to Earth-like planets in CHES, but also to systems involving massive planets, binaries, and compact objects. For stronger signals, the precision requirement can be relaxed, making the method adaptable to other astrometric missions and providing an alternative framework that reduces reliance on the \emph{Gaia} catalog.
	
	The paper is organized as follows. Section~\ref{sect:mission} introduces the CHES mission and observational data. Section~\ref{sect:model} presents the astrometric model, including proper motion, parallax, and planetary perturbations. Section~\ref{sec:Num} demonstrates the method through simulations of Earth-like planets, Jovian planets, and black holes. Finally, we discuss future improvements and remaining challenges.
	
	\section{CHES mission and simulated observation data}
	\label{sect:mission}

	Based on the CHES mission design, the satellite will operate at the Sun--Earth L2 point for a five-year observation campaign. According to the observing strategy \citep{Tan2024}, a cycle of 81 targets will be monitored, including 91 FGK-type stars within $10~\mathrm{pc}$ (including binaries). The number of observations per target ranges from 30 to 300, depending on factors such as stellar magnitude, distance, and the number of available reference stars. The observation epochs are distributed over the mission lifetime to ensure sufficient coverage of both parallax and planetary orbital periods. The designed single-measurement precision is on the order of $1~\mu\mathrm{as}$ \citep{Ji2022}.
	
	CHES employs laser interferometric calibration to achieve high-precision angular separation measurements at the level of $\sim10^{-5}$ pixels. A high-frequency stable laser produces interference fringes through heterodyne detection, which are used for separation measurements. A fiber-coupled LED source simulates stellar images for micro-pixel calibration, and CMOS detectors record both fringe and stellar signals for data processing \citep{Ji2024}. In this work, an astrometric precision of $1~\mu\mathrm{as}$ is adopted for all simulations.
	
\begin{figure}
		\centering
		\includegraphics[width=0.5\textwidth, angle=0]{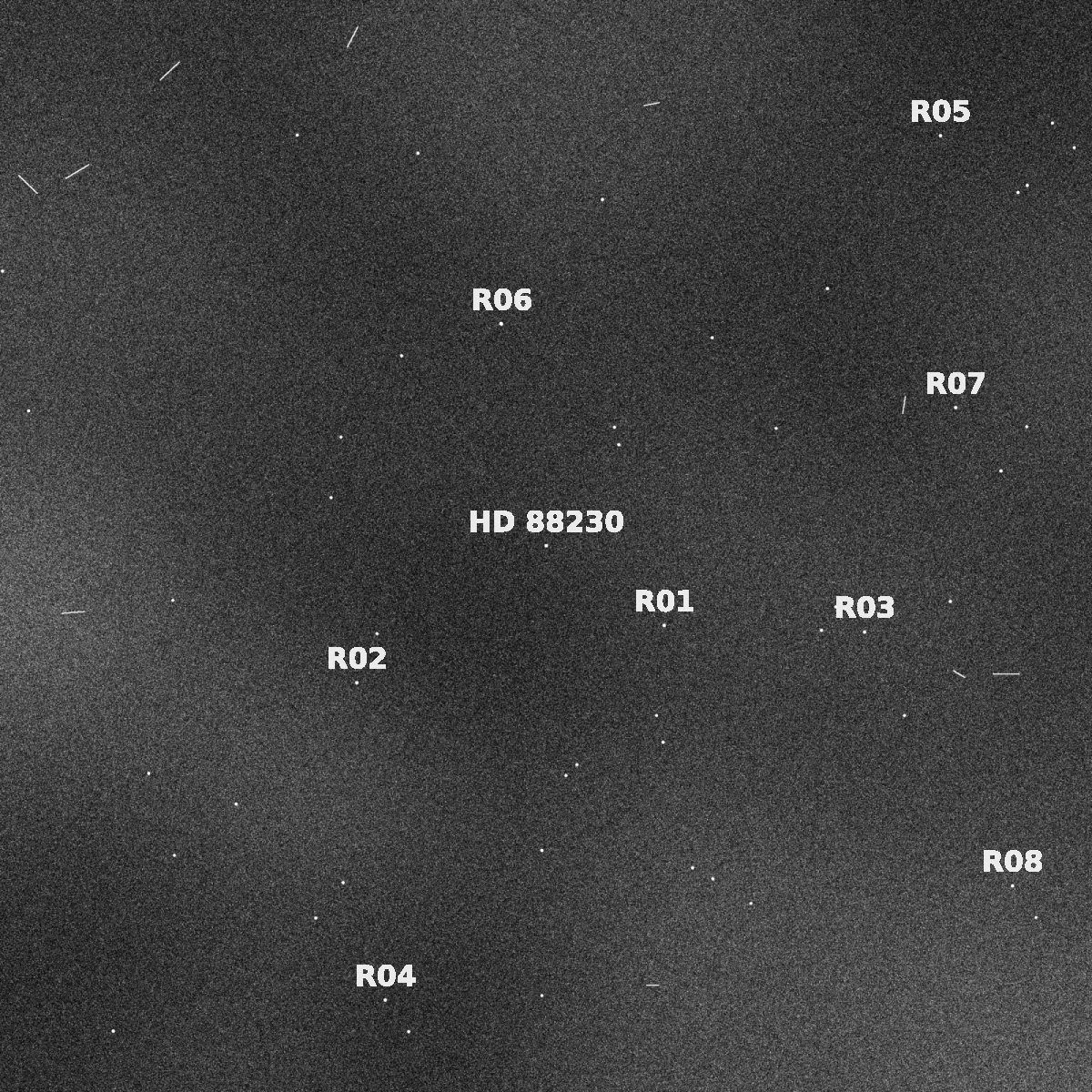}
		\caption{
			Simulated observation of HD 88230 \citep{GaiaCollaboration2023}, showing all stars brighter than magnitude 16 within a $0.44^{\circ} \times 0.44^{\circ}$ field of view. The image was downsampled from an array of 9 $\times$ 9 CMOS detectors, each with 4096 $\times$ 4096 pixels, to a final resolution of 2000 $\times$ 2000 pixels. The simulation is based on the technical specifications of the CHES optical telescope payload \citep{Ji2024}.}
		\label{distribution}
	\end{figure}

	Based on the astrometric parameters and their uncertainties from \emph{Gaia} \citep{GaiaCollaboration2023}, we simulate the observed stellar field, as shown in Figure~\ref{distribution}. Only the selected reference stars within the $0.44^{\circ} \times 0.44^{\circ}$ field of view are displayed, while fainter background stars are omitted. Reference stars are chosen from the \emph{Gaia} catalog based on magnitude, typically brighter than magnitude 13.
	
	In a narrow field of view, the prior positions of both target and reference stars are obtained from existing catalogs. However, the resulting reference frame cannot provide right ascension and declination (RA/Dec) coordinates more accurately than the catalog itself. The \emph{Gaia} DR3 catalog achieves a precision of $\sim20~\mu\mathrm{as}$ \citep{Lindegren2021}, and although future releases are expected to improve this \citep{GaiaCollaboration2023}, the temporal propagation of proper motion and parallax uncertainties leads to increasing errors, limiting its applicability for microarcsecond-level exoplanet detection.
	
	In traditional narrow-field relative astrometry (e.g., NEAT/Theia; \citealt{Malbet2012}), the observable is the differential position of the target relative to reference stars. These measurements are projected onto RA and Dec directions, requiring not only highly accurate angular separations but also precise knowledge of the observation geometry. The associated projection introduces additional uncertainties related to spacecraft attitude, catalog errors, and instrumental effects.
	
	Even without explicitly adopting a catalog-based reference frame, measuring positional changes along fixed celestial directions still requires accurate projection angles. The determination of these angles introduces further uncertainties from catalog errors, proper motion and parallax propagation, spacecraft attitude, and instrumental systematics, which directly affect the achievable precision.
	
	The geometric constraint of projection further limits performance. For example, for a separation of $0.11^{\circ}$, achieving a precision of $1~\mu\mathrm{as}$ along a given coordinate requires angular knowledge at the level of $\sim521~\mu\mathrm{as}$, excluding measurement noise and systematics. In practice, achieving such precision in spacecraft attitude is highly challenging, and reliance on reference stars ultimately reintroduces catalog limitations.
	
	Therefore, narrow-field relative astrometry has an intrinsic limitation: while angular separations can be measured with high precision, converting them into two-dimensional RA/Dec displacements requires a reference direction whose uncertainty can dominate the error budget.
	
	In contrast, the angular separation between the target and reference stars can be directly measured as a one-dimensional quantity with high precision, largely independent of catalog errors and spacecraft attitude. These separations are influenced by proper motion, parallax, binarity, and planetary perturbations, and multiple reference stars provide independent measurements for analysis.
	
	Motivated by this, we adopt an approach that avoids reliance on external reference frames and directional projections. Instead, we analyze the time variation of angular separation to extract the effects of proper motion, parallax, and planetary perturbations. In this study, stellar motions are simulated using \texttt{PyMsOfa} \citep{Ji2023}, based on the SOFA standards \citep{Hohenkerk2011}. Parallax is modeled following the formulation in Appendix~\ref{A1}, and planetary perturbations are described using the Thiele--Innes formalism \citep{Thiele1883}. These components are combined to generate simulated angular separation time series, which are then used for parameter recovery and analysis.

	\section{The fitting of angular separation length}
	\label{sect:model}
	Over the observation period, the positions of the target star and reference stars will change over time due to their proper motion, parallax, and planetary perturbations. The variation in angular separation is jointly determined by the motions of both stars. As a result, processing a set of angular separation measurements cannot separately resolve the individual parameters of the two stars. The final parameters obtained will reflect a coupled result of their combined motions.
	
	\subsection{Proper motion model}
	For the one-dimensional angular separation measurements, it is not possible to accurately derive the proper motion components in right ascension and declination. Therefore, in the proper motion fitting, we consider the total proper motion of the star on the celestial sphere. In the International Celestial Reference Frame, when the deflection of stellar trajectories due to the gravitational influence of massive objects such as binary systems and black holes is neglected, stars are assumed to move along a fixed direction, referred to as the total proper motion. Meanwhile, due to their radial velocities, a perspective acceleration arises, causing a variation in the magnitude of the total proper motion. This total proper motion and the perspective acceleration are used to compute their effects on the angular separation.
	
\begin{figure}
		\centering
		\includegraphics[width=0.7\textwidth, angle=0]{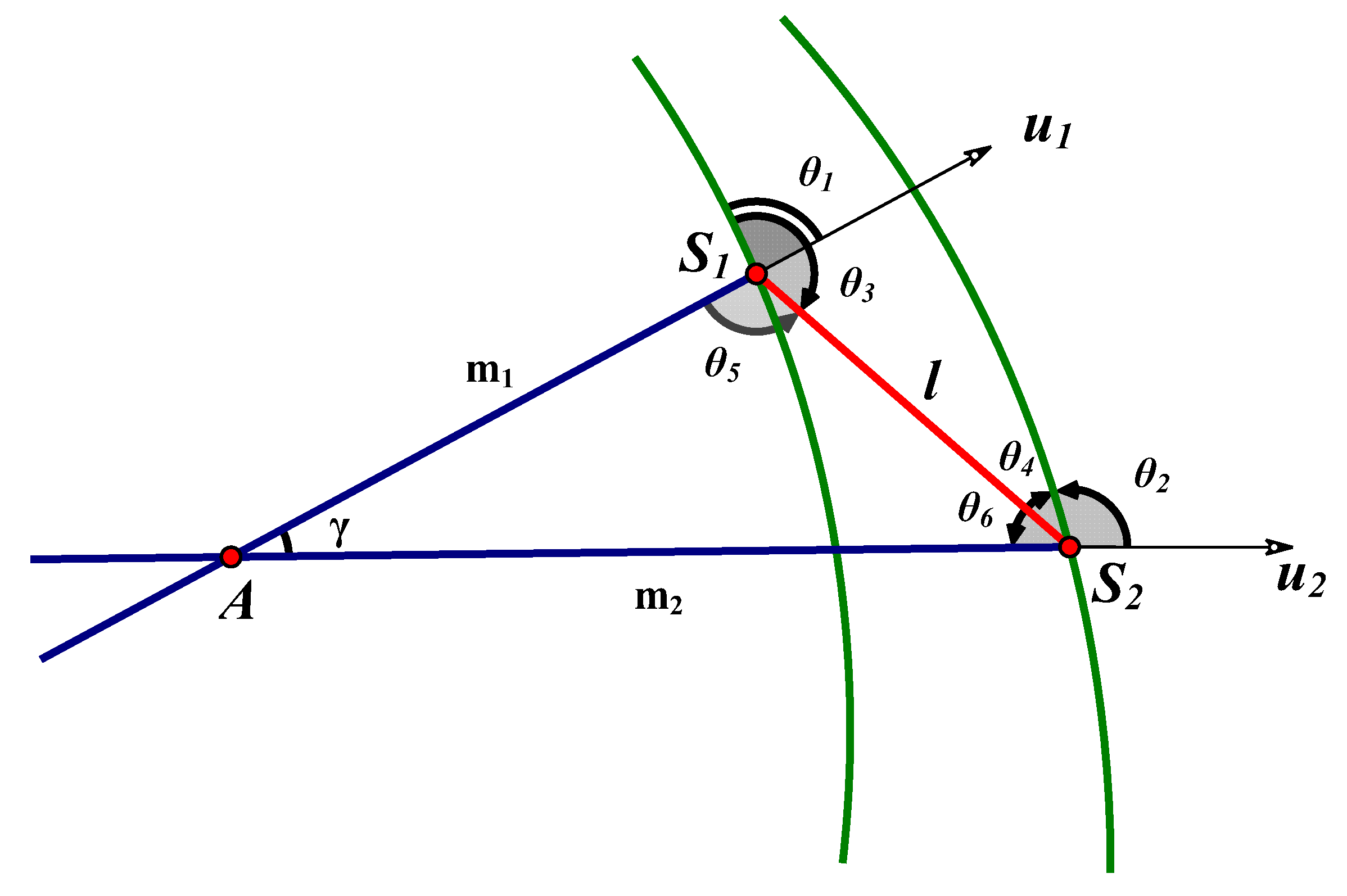}
		\caption{
			The angular separation variation caused by proper motion. $S_1, S_2$ represent two stars, $\mathbf{u_1}, \mathbf{u_2}$ denote their respective proper motion directions, and $l$ is the angular separation between the two stars.  As the stars move along their proper motion directions, the angular separation $l$ changes accordingly.}
		\label{proper motion}
	\end{figure}

	Starting from the positions of the stars ($S_1, S_2$) on the celestial sphere, we construct great circles in the direction of their total proper motion ($\mathbf{u_1}, \mathbf{u_2}$). The great circles of the two stars intersect at two points on the celestial sphere. Among these, we select the point closer to both stars as the reference point $A$, as illustrated in Figure~\ref{proper motion}. The difference in the direction of stellar proper motion will cause the star to move either closer to or farther from point $A$. The line connecting $S_1$ and $S_2$ represents the angular separation $l$.
	
	Together with a reference point $A$, the two stars ($S_1, S_2$) form a spherical triangle. Over the observation period, the stars exhibit proper motion and perspective acceleration, moving along their respective great circles in the direction of their total proper motion ($\mathbf{u_1}, \mathbf{u_2}$). The angular separation between each star and the reference point  $A$  changes at a rate determined by the total proper motion, expressed as:
	\begin{subequations}\label{c_as}
		\begin{align}
			AS_1 &= m_1 + u_1 t + a_{\mu 1} t^2,\\
			AS_2 &= m_2 + u_2 t + a_{\mu 2} t^2.
		\end{align}
	\end{subequations}
	where $m_1, m_2$ represent the initial angular separations between the stars and the reference point, respectively. $a_{\mu 1}$ and $a_{\mu 2}$ denote the perspective accelerations.
	
	Perspective acceleration is a geometric effect that arises when a star has a non-zero radial velocity, causing its observed transverse velocity (proper motion) to vary non-linearly over time. The value of perspective acceleration $a_\mu$ is determined by the star's radial velocity $V_r$, parallax $\varpi$, and proper motion $\mu$, and is calculated using the following formula:
	\begin{eqnarray}
		a_\mu = -2 \frac{\mu \times V_r}{\varpi}.
		\label{c_ua}
	\end{eqnarray}
	
	In this process, the stellar parallax also exhibits slight variations due to the presence of radial velocity, which cannot be neglected in high-precision measurements. Therefore, its variation must also be taken into account during the fitting process. When the proper motion direction moves away from the reference point $A$, the proper motion rate $u$ is defined as positive; conversely, when the motion direction approaches $A$, $u$ is considered negative. The sign of the perspective acceleration is determined jointly by the direction of the proper motion and the radial velocity. Throughout the proper motion of the stars, the reference point $A$ remains fixed, ensuring that the angle $\gamma$ between the two proper motion directions remains unchanged. In the resulting spherical triangle ($S_{1}AS_{2}$), the cosine rule can be applied as follows:
	\begin{equation}
		\begin{array}{c}
			\cos(l_{pm}(t)) = \cos AS_1\cos AS_2 + \sin AS_1 \sin AS_2 \cos \gamma \\
			= \cos(m_1 + u_1 t + a_1 t^2)\cos(m_1 + u_2 t + a_2 t^2)\\
			+ \sin(m_1 + u_1 t + a_1 t^2)\sin(m_1 + u_2 t + a_2 t^2)\cos \gamma.
		\end{array}
	\end{equation}
	where $l_{pm}(t)$ represents the influence of the proper motion component on the angular separation variation. Based on this formula, the proper motion term in the angular separation variation can be fitted using 9 parameters ($m_1, m_2, u_1, u_2, V_{r1}, V_{r2}, \gamma, \varpi_1,\varpi_2$).
	
	Using the \texttt{PyMsOfa} package \citep{Ji2023}, we simulated the motion of the target star HD 88230 and the brightest reference star (Gaia DR3 823765969936234368) \citep{GaiaCollaboration2020} on the celestial sphere, as well as the variation in angular separation between them. In \texttt{PyMsOfa}, this simulation is based on calculating the positional changes of stars in three-dimensional space due to their motion and then projecting these positions onto the celestial sphere to simulate their post-proper-motion locations. We then calculate the angular separation between the target star and the reference star based on their positions.
	
	Using the parameters of the target star and the reference star adopted in the simulation, the angular separation was computed based on the aforementioned model, achieving a precision of $0.02~\mu \rm as$. The residual error originates from the mutual coupling between the proper motion and the perspective acceleration, as the proper motion is influenced by the perspective acceleration, while the computation of the latter also depends on the proper motion.
	
	\subsection{Parallax model}
\begin{figure}
		\centering
		\includegraphics[width=0.52\textwidth]{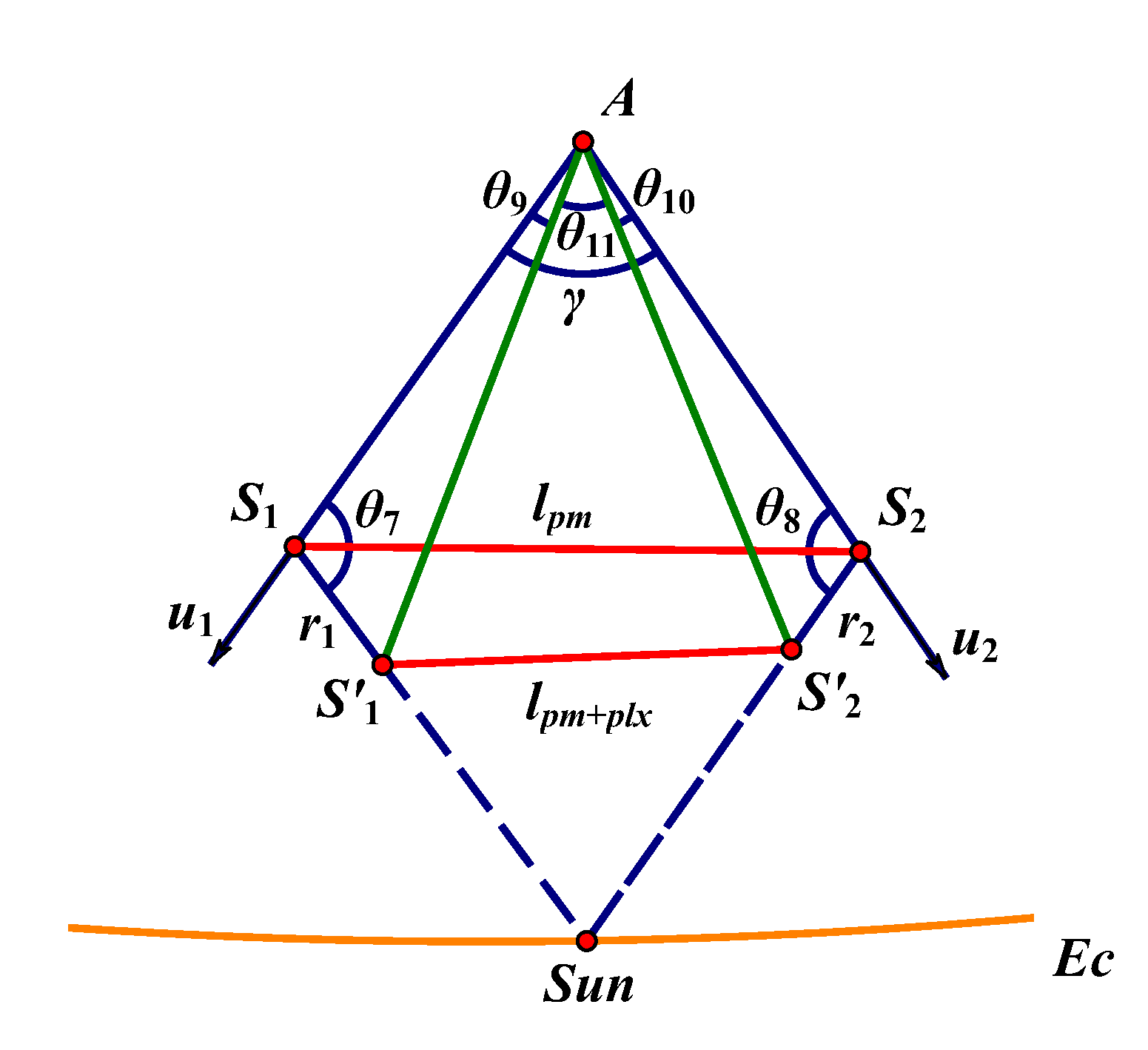}
		\includegraphics[width=0.45\textwidth]{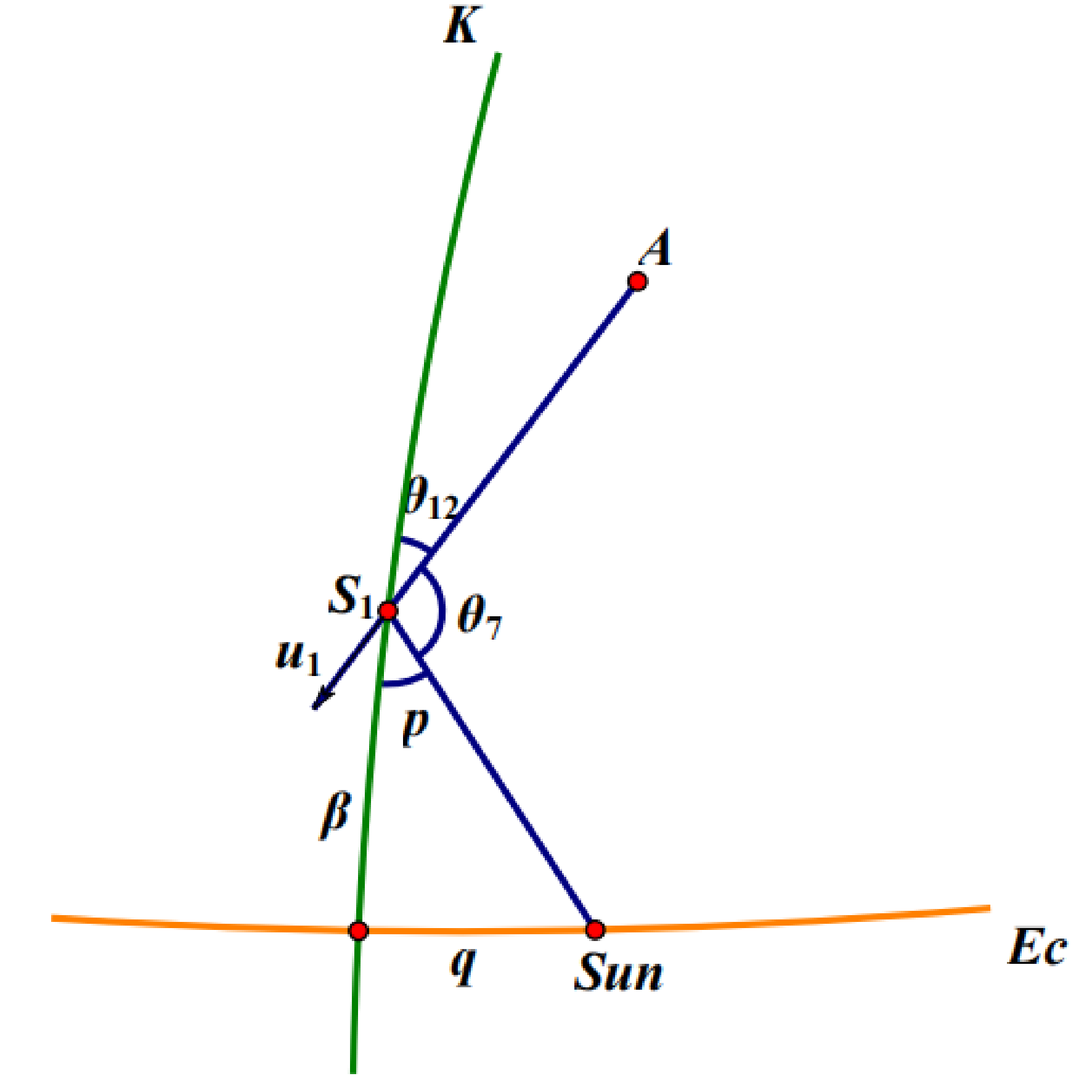}
		\caption{
			Parallax is added to the angular separation variation caused by proper motion.
			\emph{Left panel}: $l_{\rm pm}$ denotes the angular separation considering only proper motion; parallax shifts the stellar positions toward the Sun by distances $r_1$ and $r_2$, resulting in a new separation $l_{\rm pm+plx}$.
			\emph{Right panel}: Geometry of the relevant angles in the model, defined using the stars' ecliptic longitude and latitude.
		}
		\label{plx0}
	\end{figure}

	Parallax refers to the annual apparent shift in the position of a star on the celestial sphere caused by the observer's orbit around the Sun. For the CHES satellite observing from the L2 point, the observed parallax of a star will be about 1.01 times the parallax value defined in the ICRS. Under the effect of parallax, the position of a star on the celestial sphere will shift toward the direction of the Sun. The distance of this displacement ($r$) depends on the distance between the star and the observer, the relative position between the star and the Sun, and the star's ecliptic latitude:
	\begin{equation}
		\begin{split}
			r &= \varpi
			\sqrt{\frac{\sin^2 \beta}
				{\cos^2 p + \sin^2 \beta \sin^2 p}},\\[6pt]
			\cot p &= \sin \beta \, \cot q,
		\end{split}
		\label{cr}
	\end{equation}
	where $r$ denotes the displacement distance, $\varpi$ represents the parallax of the star and varies with the radial velocity over the course of the observation period. $\beta$ is its ecliptic latitude, and $p$ is the angle between the line connecting the star and the Sun and the direction of ecliptic longitude, $q$ represents the difference in ecliptic longitude between the sun and the star, as shown in Figure~\ref{plx0}(b).
	
	Building upon the proper motion model presented in the previous section, we now incorporate the effect of parallax on stellar positions, as illustrated in Figure~\ref{plx0}(a). In the ecliptic coordinate system, two stars within the field of view ($S_1$ and $S_2$) move along their respective proper motion directions ($\mathbf{u}_1$ and $\mathbf{u}_2$). Under the influence of parallax, they are each displaced by a certain distance ($r_1$ and $r_2$) toward the direction of the Sun. When only proper motion is considered, their angular separation is represented by \textsc{$l_{pm}$} in the figure. When parallax is included, the angular separation becomes $l_{pm+plx}$, as shown.
	
	Among the parameters shown in the figure, $AS_1$ and $AS_2$ are given by Equation~\ref{c_as} from the previous section. $r_1$ and $r_2$ are calculated using the parallaxes of the two stars and the parallax factors via Equation~\ref{cr}. $\theta_7$ is defined in Figure~\ref{plx0}(b) and are given by:
	\begin{equation}
		\theta_7 = \pi - p - \theta_{12},
		\label{c_theta78}
	\end{equation}
	where $\theta_{12}$ represents the angle between the direction of the star's proper motion and the ecliptic longitude. Since this angle varies with the proper motion, both its initial value ($w_1$) and its rate of change ($uw_1$) are treated as fitting parameters in this model. The value of $\theta_8$ is calculated in a similar manner with $w_2$ and $uw_2$.
	
	Based on $\theta_7$ and $\theta_8$, we can use spherical trigonometric formulas to calculate $\theta_9$ and $\theta_{10}$, as well as $AS'_1$ and $AS'_2$:
	\begin{equation}
		\begin{split}
			\cot \theta_9 &=
			\frac{1}{\sin \theta_7}
			\left( \sin AS_1 \cot r_1 - \cos AS_1 \cos \theta_7 \right),\\[6pt]
			\cot \theta_{10} &=
			\frac{1}{\sin \theta_8}
			\left( \sin AS_2 \cot r_2 - \cos AS_2 \cos \theta_8 \right),\\[6pt]
			\cos AS'_1 &=
			\cos AS_1 \cos r_1 + \sin AS_1 \sin r_1 \cos \theta_7,\\[6pt]
			\cos AS'_2 &=
			\cos AS_2 \cos r_2 + \sin AS_2 \sin r_2 \cos \theta_8.
		\end{split}
		\label{c_theta910}
	\end{equation}
	
	Using the fixed angle $\gamma$ between the proper motion directions of the two stars, as defined in the proper motion model, $\theta_{11}$ can be calculated by:
	\begin{equation}
		\theta_{11} = \gamma + \text{sign}_1 \theta_9 + \text{sign}_2 \theta_{10},
		\label{c_theta11}
	\end{equation}
	where, the value of $\text{sign}_1$ and $\text{sign}_2$ are either $+1$ or $-1$, determined by the geometric relationship among the positions of the Sun, the stars, and the intersection point ($A$) of the stars' proper motion direction. This value can be predicted using prior parameters.
	
	The value of $l_{pm + plx}$ is given by the law of cosines:
	\begin{equation}
		\begin{split}
			\cos l_{pm + plx}
			&= \cos AS'_1 \cos AS'_2\\
			&\quad + \sin AS'_1 \sin AS'_2 \cos \theta_{11}.
		\end{split}
		\label{c_lpmplx}
	\end{equation}
	
	Combining Equations~\ref{c_as} and \ref{cr}-\ref{c_lpmplx} yields our one-dimensional angular separation variation model that incorporates both proper motion and parallax.
	
	Based on the parallax principle, we simulated the positions of the target and reference stars within the field of view as observed by a satellite located at the L2 point, thereby obtaining the variation in angular separation under the influence of proper motion and parallax. Using the stellar parameters adopted in the simulation, the real values of the 13 parameters ($m_1, m_2, u_1, u_2, V_{r1}, V_{r2}, \gamma, \varpi_1,\varpi_2,w_1, w_2, uw_1, uw_2$) described in the model were also calculated. The difference between the angular separations computed from the model and those obtained from the simulation does not exceed $0.1~\mu \rm as$.
	
	\subsection{Aberration and Gravitational Lensing}
	
	In addition to positional variations caused by proper motion and parallax, the apparent positions of stars observed by space-based astrometric satellites are affected by several physical effects. Among them, stellar aberration and gravitational lensing are the dominant contributors that must be modeled in high-precision astrometry \citep{Klioner2003, Soffel2003}.
	
	Stellar aberration arises from the motion of the observer relative to incoming starlight, causing a displacement of the apparent position toward the direction of motion. For a satellite near the Sun-Earth L2 point, this effect can reach several tens of milliarcseconds \citep{Klioner2003}. Gravitational lensing results from light deflection by massive bodies in the solar system, dominated by the Sun \citep{Soffel2003}. In the CHES observing strategy, the target-Sun separation is kept above $60^{\circ}$, limiting the deflection to several tens of microarcseconds, while other solar system bodies introduce additional microarcsecond-level perturbations \citep{Klioner2003}.
	
	These effects are mainly determined by the satellite motion and external gravitational fields, and are therefore nearly identical for stars within the same field of view, with only small differential contributions.
	
	To correct for these effects, we compute the corresponding correction factors using prior stellar positions, satellite kinematics, and solar system ephemerides. Based on \emph{Gaia} astrometric parameters, we derive the expected stellar positions at each epoch and calculate the angular separation corrections due to gravitational lensing ($k_{gl}$) and stellar aberration ($k_{ab}$). The corrected angular separation is given by
	\begin{equation}
		l_{mod} \approx k_{ab}(t),k_{gl}(t),l_{L2},
	\end{equation}
	where $l_{mod}$ is the observed separation and $l_{L2}$ includes only proper motion and parallax.
	
	The satellite position and velocity are simulated using the orbital model described in Appendix~\ref{app:orbit}, and used to compute time-dependent corrections. After applying these corrections, the residuals in angular separation are reduced to $\sim10^{-3}~\mu$as, demonstrating that the systematic effects are effectively removed.

	\subsection{Planet perturbation model}
	The fitting of planetary motion often utilizes the methods derived from the TI (Thiele-Innes) formula \citep{Thiele1883}, where the fitting is conducted within the observational plane to account for the positional shifts of a star caused by planetary motion. However, when only the variation in angular separation is known, it is impossible to directly determine the planet's specific motion on the celestial sphere; therefore, the model presented in this paper is not applicable for direct fitting.
	\begin{figure}
		\centering
		\includegraphics[width=0.45\textwidth, angle=0]{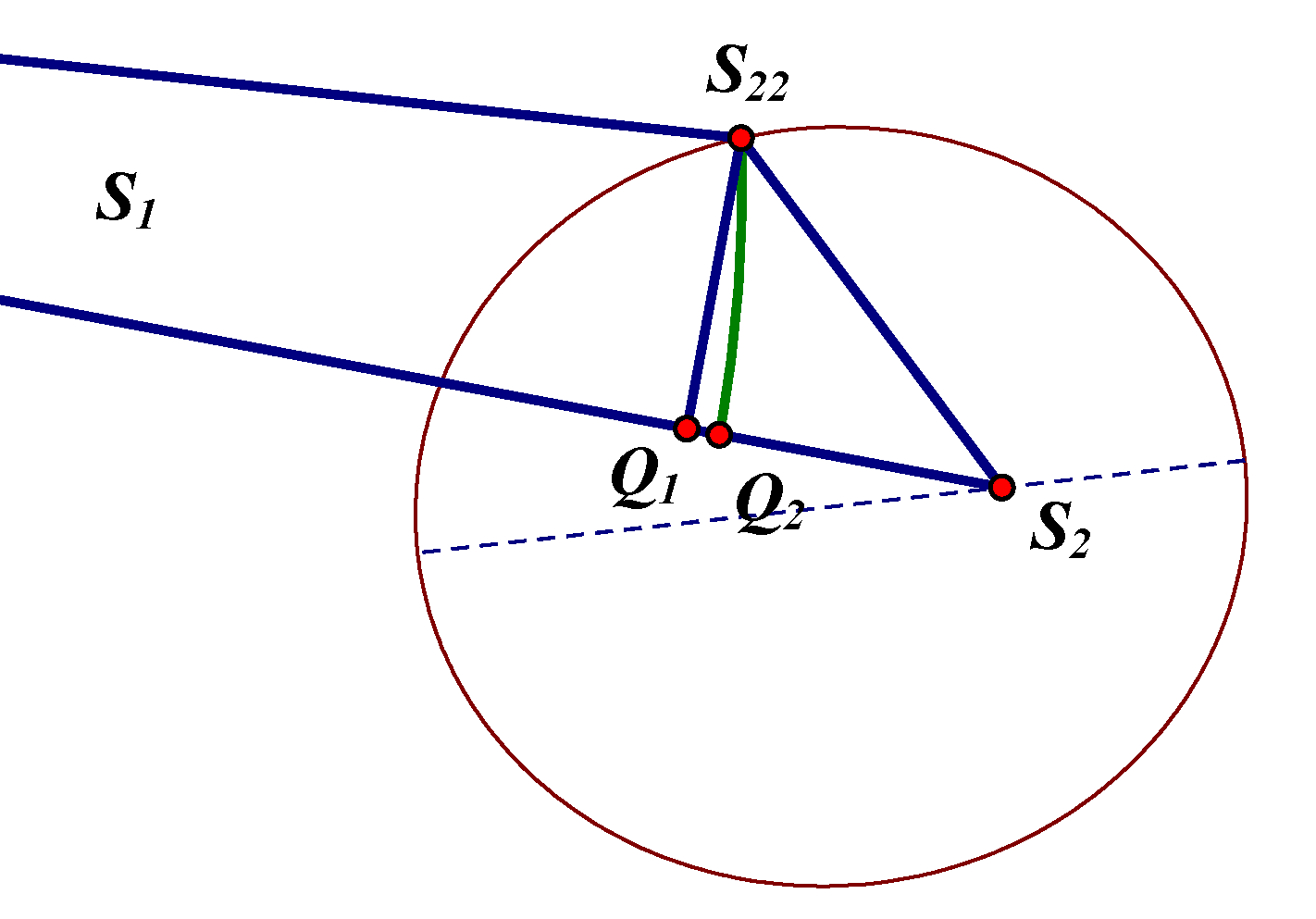}
		\caption{
			The star $S_2$, influenced by gravitational perturbations from its planet, shifts its position to point $S_{22}$. Its projection in the angular separation direction is represented as point $Q_1$ in the diagram. $S_1Q_2$ illustrates the actual length of $S_1S_{22}$ in the angular separation direction.
		}
		\label{planet1}
	\end{figure}
	
	We first assume that only one planet exists around the target star. In this case, the angular separation variation induced by the planet affects only one end of the separation. The position of that end is projected onto the direction of the angular separation unaffected by the planetary perturbation, as illustrated in Figure~\ref{planet1}. The target star's projected position along the angular separation is denoted as $Q_1$. The difference in length between $S_1Q_1$ and $S_1S_{22}$ is represented as $Q_1Q_2$, which is on the order of $ 10^{-4}~\rm \mu \rm as $ for an Earth-like planet located in the habitable zone. Thus, we substitute the fitting of the angular separation variation caused by the planet with the fitting of $S_1Q_1$'s length.
	
	In the orbital plane, considering Kepler's laws, the position vector can be expressed as a two-dimensional vector:
	
	\begin{equation}
		\mathbf{r_1} = \begin{pmatrix} x\\y\end{pmatrix} = a\begin{pmatrix} \cos E-e\\\sqrt{1-e^2}\sin E \end{pmatrix}.
	\end{equation}
	where $E$ represents the eccentric anomaly, $e$ is the orbital eccentricity, and $a$ is the semi-major axis of the orbit. We represent the position of the star in the orbital plane by a set of orthogonal basis vectors $\mathbf{u}$ and $\mathbf{v}$, such that the position vector is expressed as $\mathbf{r_1} = x\,\mathbf{u} + y\,\mathbf{v}$. Let $\mathbf{n}$ denote the unit vector defining the direction of the angular separation in three-dimensional space. The projection of the stellar position from the orbital plane onto the direction of angular separation can then be written as:
	\begin{equation}
		l_{pl3D} = a[(\cos E - e)(\mathbf{u} \cdot \mathbf{n})	+ (\sqrt{1-e^2}\sin E)(\mathbf{v} \cdot \mathbf{n}) ].
	\end{equation}
	
	When projecting its effect on the angular separation onto the observation plane and simplifying the formula, it becomes:
	
	\begin{equation}
		l_{pl} = D_1 (\cos E - e) + D_2 (\sqrt{1-e^2}\sin E).
		\label{pl}
	\end{equation}
	
	Here, $D_1$ and $D_2$ encapsulate the spatial positional relationship between the orbital plane and the line connecting the two stars, as well as the projection relationship from space onto the observation plane. The fitted values of $D_1$ and $D_2$ will not exceed the semi-major axis $a$ of the orbit. The calculation of the eccentric anomaly $E$ involves Kepler's equation, given as:
	
	\begin{equation}
		E - e \sin E = \frac{2 \pi}{P}t - M_0.
		\label{kepler}
	\end{equation}
	Therefore, the fitting parameters will also include the orbital period  $P$  and the mean anomaly  $M_0$  at $t = 0$.
	
	We simulated the orbital motion of a star hosting a single planet using the TI formula and calculated the resulting variation in angular separation caused solely by the planetary perturbation. This variation was then used as the observational input for the planetary component fitting. The difference between the angular separations derived from the fitted parameters and those from the simulation is less than $0.005~\mu \rm as$. However, since the fitted quantity represents only the projection of the stellar motion along the direction of angular separation, it is impossible to determine the complete set of orbital parameters based solely on the angular separation with a single reference star. Only the eccentricity $e$, orbital period $P$. Moreover, due to the one-dimensional nature of the angular separation, the direction of the stellar motion cannot be determined from the mean anomaly $M_0$, which therefore must be constrained within the range of $0 - \pi$. In addition, when the strength of the planetary signal is comparable to the observational uncertainty of the angular separation, the fitted parameters of the planetary model will exhibit significant errors, which will be discussed in detail later.
	
	In our observations, the field of view typically contains no fewer than six reference stars. During the observation process, the angular separations between the target star and all reference stars within the field can be measured simultaneously. The angular separations corresponding to reference stars in different directions provide projections of the stellar motion induced by planetary perturbations along multiple directions, enabling the reconstruction of the complete stellar motion within the observational plane. During this two-dimensional reconstruction process, the prior position of the star can be used as a reference for the direction of motion projection, and the positional uncertainty of the star does not significantly affect the accuracy of the reconstructed trajectory. Consequently, the stellar motion induced by the planetary perturbation can be reconstructed in the equatorial coordinate system based on the fitting results, from which other relevant parameters can be derived using the TI formula or similar methods. The detailed procedure for fitting the planetary orbital parameters is beyond the scope of this paper and will not be discussed further here. Numerous studies have addressed this topic, and in our analysis we adopt the fitting approach described in \citet{Bao2024b}, which provides a comprehensive framework for deriving planetary orbits from astrometric signals.
	
	\subsection{Angle Distance Fitting}
		
	In the preceding four subsections, we discussed the effects of proper motion, parallax, aberration of light, gravitational lensing, and planetary perturbations on the variation of angular separation. By combining these effects, the total variation in angular separation between the target and reference stars can be expressed as
	
	\begin{align}
		l_{\rm mod}(t)
		&= k_{\rm gl}(t)\, k_{\rm ab}(t)\,
		f\!\left(l_0,
		m_i, \mu_i, \gamma,
		V_{ri}, \varpi_i, w_i, uw_i,
		\right.
		\notag \\[-4pt]
		&\qquad\qquad\qquad\left.
		e,\, P,\, M_0,\, D_1,\, D_2,\,t
		\right),\quad (i=1,2)
		\label{eq:lobs}
	\end{align}
	
	Here, the proper-motion model depends on $m_i, \mu_i, \gamma, V_{ri}, \varpi_i$; the parallax model uses $V_{ri}, \varpi_i, w_i, uw_i$; and the planetary perturbation model employs $e, P, M_0, D_1, D_2$.
	
	Besides the fitted parameters, the prior ecliptic coordinates of the target star and the Sun are required as auxiliary inputs. The complete model is summarized in Appendix~\ref{app:model_summary}.
	
	In total, the model contains 18 linearly independent parameters, all determined from one-dimensional measurements of angular separation variations. Table~\ref{tab:fit_params} lists these parameters along with their physical meanings. Parameter estimation is performed using the \texttt{dynesty} nested sampling algorithm, which efficiently explores the posterior distribution and naturally evaluates Bayesian evidence. This approach is well-suited for models combining linear and nonlinear components.
	
	The likelihood function assumes Gaussian measurement uncertainties in the angular separation:
	
	\begin{equation}
		\mathcal{L} =
		\exp\left[
		-\frac{1}{2} \sum_{t}
		\frac{\bigl(l_{\rm ob}(t) - l_{\rm mod}(t)\bigr)^2}{\sigma_l^2}
		\right],
		\label{eq:likelihood}
	\end{equation}
	
	where $\sigma_l$ is the measurement uncertainty. Together with uniform or catalogue-based priors, this likelihood forms the basis for Bayesian inference of all fitted parameters.
	
	\begin{table*}[htbp]
		\centering
		\caption{List of fitted parameters included in the model.}
		\label{tab:fit_params}
		\small
		\setlength{\tabcolsep}{6pt}
		\begin{tabular}{ll}
			\hline
			{Parameter} & {Physical meaning} \\
			\hline
			$m_1$, $m_2$ & Initial distance between the star and the intersection point along the proper motion direction.\\
			$\mu_1$, $\mu_2$ & Magnitude of total proper motion.\\
			$\gamma$ & Position angle of proper motion at the intersection point.\\
			$V_{r1}$, $V_{r2}$ & Radial velocity.\\
			$\varpi_1$, $\varpi_2$ & Stellar parallax.\\
			$w_1$, $w_2$, $uw_1$, $uw_2$ & Angle between the proper motion direction and ecliptic longitude, and its rate of change.\\
			$e$, $P$, $M_0$ & Orbital eccentricity, period, and mean anomaly of the planetary perturbation.\\
			$D_1$, $D_2$ & Projection coefficient of orbital motion along the angular separation direction.\\
			\hline
		\end{tabular}
	\end{table*}
	
	All parameters except the five planetary ones can be assigned priors from \emph{Gaia} or other astrometric catalogs. For stars with small total proper motion, the change in angle between the proper motion direction and the ecliptic longitude can be neglected, reducing the number of fitted parameters and improving computational efficiency.
	
	We evaluate the model under different signal-to-noise ratios (SNR), assuming an angular separation measurement uncertainty of $\sim 1~\mu\mathrm{as}$. Planetary-induced stellar reflex motions vary widely: Earth-like planets in habitable zones produce signals of $\sim 1~\mu\mathrm{as}$, warm Jupiters several tens of $\mu\mathrm{as}$, and Jupiter analogs several hundreds of $\mu\mathrm{as}$. Consequently, Jupiter analogs correspond to SNRs of several hundred, and warm Jupiters to SNRs of several tens. In these high-SNR regimes, the planetary perturbation dominates the measurement noise, allowing robust fitting of all five planetary parameters and high-fidelity reconstruction of the full two-dimensional orbital motion.
	
	For Earth-like planets with amplitudes comparable to the noise, the SNR approaches unity, making direct fitting of all planetary parameters unreliable. In this low-SNR regime, the thirteen non-planetary parameters are first determined without including the planetary perturbation. The residuals between observed and model-predicted separations, which contain both planetary signals and observational noise, are then used to reconstruct the orbit. Although this residual-based approach reduces precision and increases uncertainty compared to high-SNR fitting, it retains the coherent orbital information and remains effective in marginal SNR cases where direct multi-parameter fitting becomes unstable.
	
	\begin{table*}
		\centering
		\caption{Fit parameters with uncertainties and true values.}
		\label{fit-para}
		\begin{tabular}{lccc}
			\hline
			Parameter (unit) & Fit value & Uncertainty & True value \\
			\hline
			$m_1$ (rad)      & $1.514\times10^{-2}$			& $6.55\times10^{-8}$ 	& $1.515\times10^{-2}$ 	\\
			$m_2$ (rad)      & $1.328\times10^{-2}$ 		& $6.92\times10^{-8}$ 	& $1.329\times10^{-2}$ 	\\
			$\mu_1$ (rad\,yr$^{-1}$) & $4.948\times10^{-6}$ & $4.82\times10^{-12}$ 	& $4.948\times10^{-6}$ 	\\
			$\mu_2$ (rad\,yr$^{-1}$) & $3.349\times10^{-7}$ & $1.73\times10^{-11}$ 	& $3.351\times10^{-7}$ 	\\
			$\gamma$ (rad)   & $1.755\times10^{-1}$ 		& $6.76\times10^{-7}$ 	& $1.753\times10^{-1}$ 	\\
			$V_{r1}$ (km\,s$^{-1}$) & $-2.634\times10^{1}$ 	& $1.41\times10^{-2}$ 	& $-2.648\times10^{1}$	\\
			$V_{r2}$ (km\,s$^{-1}$) & $3.247\times10^{1}$ 	& $2.24\times10^{-3}$ 	& $3.273\times10^{1}$ 	\\
			$\varpi_1$ (arcsec) & $2.053\times10^{-1}$ 		& $9.28\times10^{-7}$ 	& $2.053\times10^{-1}$ 	\\
			$\varpi_2$ (arcsec) & $1.082\times10^{-2}$ 		& $6.82\times10^{-7}$ 	& $1.081\times10^{-2}$ 	\\
			$w_1$ (rad)      & $5.684\times10^{0}$			& $3.77\times10^{-5}$ 	& $5.680\times10^{0}$ 	\\
			$w_2$ (rad)      & $5.940\times10^{0}$			& $7.50\times10^{-4}$ 	& $5.857\times10^{0}$ 	\\
			$u w_1$ (rad\,yr$^{-1}$) & $2.906\times10^{-6}$	& $7.33\times10^{-7}$ 	& $2.045\times10^{-6}$ \\
			$e$           & $3.138\times10^{-1}$			& $2.93\times10^{-3}$ 	& $3.000\times10^{-1}$ \\
			$T$ (yr)         & $3.876\times10^{-2}$ 		& $4.59\times10^{-7}$ 	& $3.876\times10^{-2}$ \\
			$M_0$ (rad)      & $9.068\times10^{-1}$ 		& $1.09\times10^{-2}$ 	& $7.854\times10^{-1}$ \\
			$D_1$ ($\mu as$) & $2.529\times10^{1}$			& $4.97\times10^{-2}$ 	& - \\
			$D_2$ ($\mu as$) & $5.570\times10^{0}$ 			& $2.77\times10^{-1}$ 	& - \\
			\hline
		\end{tabular}
	\end{table*}
	
	As an example, in Sect.~\ref{subsec:JP} we perform a fit to 125 angular-separation measurements between the target star and one reference star in the hot-Jupiter case. Table~\ref{fit-para} reports the fitted values of all 17 parameters. The uncertainties listed in the table correspond to statistical errors derived from the posterior distributions, representing precision under the adopted model and not including potential systematic errors due to model limitations. Except for parallax and radial velocities, all fitted parameters correspond to the first observation epoch. The small rate of change of the angle between the reference star's proper-motion direction and the ecliptic longitude ($uw_2$) was omitted to improve computational efficiency. The geometric coefficients $D_1$ and $D_2$ are not assigned explicit numerical values here.
	
	With a measurement uncertainty of $1~\mu\mathrm{as}$, the hot-Jupiter perturbation has a high SNR. Therefore, the two-dimensional orbital motion can be reconstructed either (i) directly from the five planetary parameters $(e,\,T,\,M_0,\,D_1,\,D_2)$, or (ii) by using the residual angular separations obtained after subtracting the model based on the first 12 parameters, following the procedure appropriate for lower-SNR cases.
	
	For a target star observed with $n$ reference stars, $C_n^2$ unique angular separation combinations can be constructed. Each combination provides an independent constraint on the stellar motion, improving precision and reducing uncertainties in the reconstructed two-dimensional orbit. However, when the target star and two reference stars are nearly collinear, the corresponding angular separation vectors are almost aligned. Such degenerate combinations are therefore excluded to ensure stable and reliable orbital reconstruction.

	\section{Numerical Simulation Cases}
	\label{sec:Num}
	The CHES mission aims to detect and characterize Earth-like planets within the habitable zones of nearby solar-type stars, located at distances of approximately 10 pc, using microarcsecond-level relative astrometry.
 When modeling the length of the semi-major axis, the approach is based on the perturbations exerted by planets on their host stars. The description of stellar orbits in Equation~\ref{pl} is not limited to systems with exoplanets; it also applies to binaries and black holes. Consequently, this model is broadly applicable to any scenario where stellar motion adheres to Keplerian orbits.
	
	\subsection{Earth-like Planet}
	
	The wobble of the stellar motion signal induced by terrestrial planetary perturbations, typically on the microarcsecond scale, is collectively constrained by the distance between the star and its habitable-zone planet, the planetary mass required for an Earth-like planet, and the distance between the star and the observer. To assess whether the method described in the previous sections can achieve the required observational precision, we select HD 88230 as our target star, as previously mentioned. Among the 15 reference stars in its field of view, we choose 8 with magnitudes brighter than 12 as our observational targets. For this analysis, we assume the presence of an Earth-like planet within the habitable zone of the target star.
	
	Using the \texttt{PyMsOfa} package, together with the parallax principle, the TI formulation, and the relativistic corrections determined by the instantaneous position and velocity of the observing satellite (i.e., stellar aberration and gravitational light deflection), we simulated the apparent positions of the target and reference stars within the field of view at the observation epochs defined by the adopted observing strategy \citep{Tan2024}. Based on this simulation, we calculated the angular separations between the target star and the reference stars that can be obtained during actual observations. In the CHES design, the observational error is specified to be $1~\mu \rm as$. By calculating the angular separations between the two stars along the $x$ and $y$ directions on the photographic plate and applying the error propagation formula {\citep{taylor1997}}:
	
	\begin{equation}
		\begin{array}{rl}
			L =& \sqrt{x^2 + y^2} \\
			\Delta L =& \sqrt{ (\frac{\partial L}{\partial x}\Delta  x)^2+ (\frac{\partial L}{\partial y}\Delta  y)^2 } .
		\end{array}
	\end{equation}
	The measurement error for the angular separations is also $1~\mu \rm as$. Ultimately, the data used for processing consist of eight sets of time-varying angular separation information between the target star and the reference stars.
	
	After fitting with the model described in the previous sections, we removed the proper motion and parallax terms from the angular separation data and performed a two-dimensional reconstruction. The resulting data is shown as the red points in Figure~\ref{Exoplanet}. Based on this, we obtained the relevant parameters for planetary perturbations through a Keplerian fit, as listed in Table~\ref{planetfit}.
	
	\subsection{Jovian Planet}
	\label{subsec:JP}
	
	Jovian planets are widely distributed and constitute a significant component of planetary systems. Observations indicate that such planets are not only prevalent in our solar system but are also frequently detected in planetary systems \citep{Mayor1995}. More massive Jovian planets, comparable to Jupiter and Saturn, are often more readily observed due to their stronger gravitational and photometric signatures. Their masses span approximately 0.3 to several times the mass of Jupiter, and their orbital periods range from a few days (hot Jupiters) to several decades (long-period giants), reflecting diverse dynamical histories. The study of these planets provides critical insights into planetary system formation and evolution \citep{Pollack1996}.
	
	Hot Jupiters, which orbit close to their host stars, have become research focal points due to their unique characteristics. Discoveries of these planets have provided crucial evidence for planetary migration theories and revealed complex planet-star interactions \citep{Matsumura2007,Dawson2018}.
	
	The model developed here is applicable to planetary systems hosting Jovian planets. To validate its performance, we conduct additional simulations for both warm Jupiter and classical Jovian planet scenarios. As a representative warm Jupiter case, we adopt HD 88230, which hosts a planet with an orbital period of 14.16 days. If transits occur, the planet can be detected via the transit method; otherwise, astrometry becomes essential. For the Jovian planet case, HD 147513 hosts the confirmed planet HD 147513 b \citep{Mayor2004}, with its parameters listed in Table~\ref{planetfit}.
 Radial velocity measurements provide a minimum mass of $M \sin i = 1.21~\rm M_{Jup}$, while astrometry can determine the true mass. In simulating the planetary perturbation, we assume $\sin i = 0.6$ and a mass of $614.472~\rm M_{\oplus}$.
	
\begin{figure}
		\centering
		\includegraphics[width=0.32\textwidth]{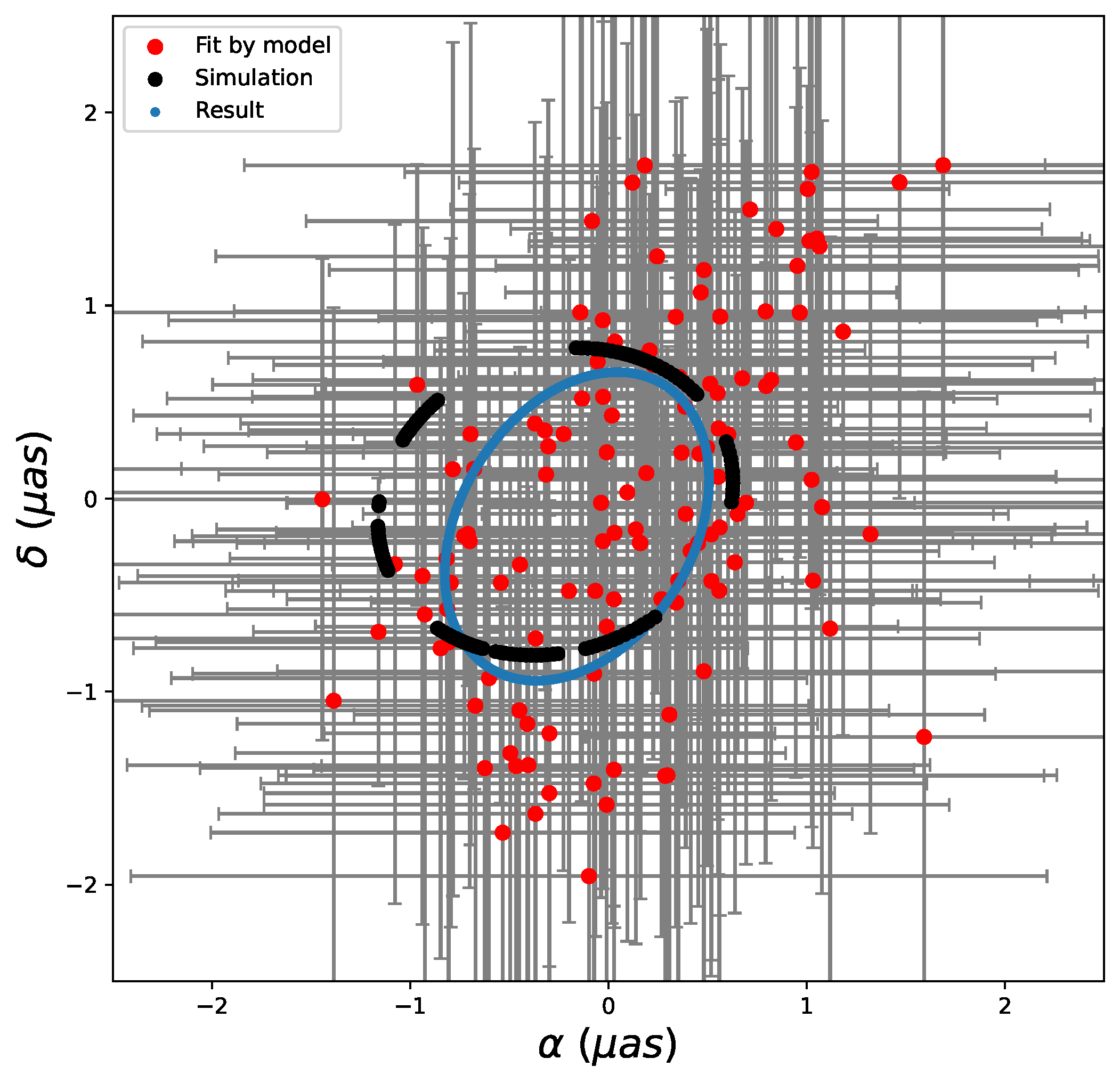}
		\includegraphics[width=0.32\textwidth]{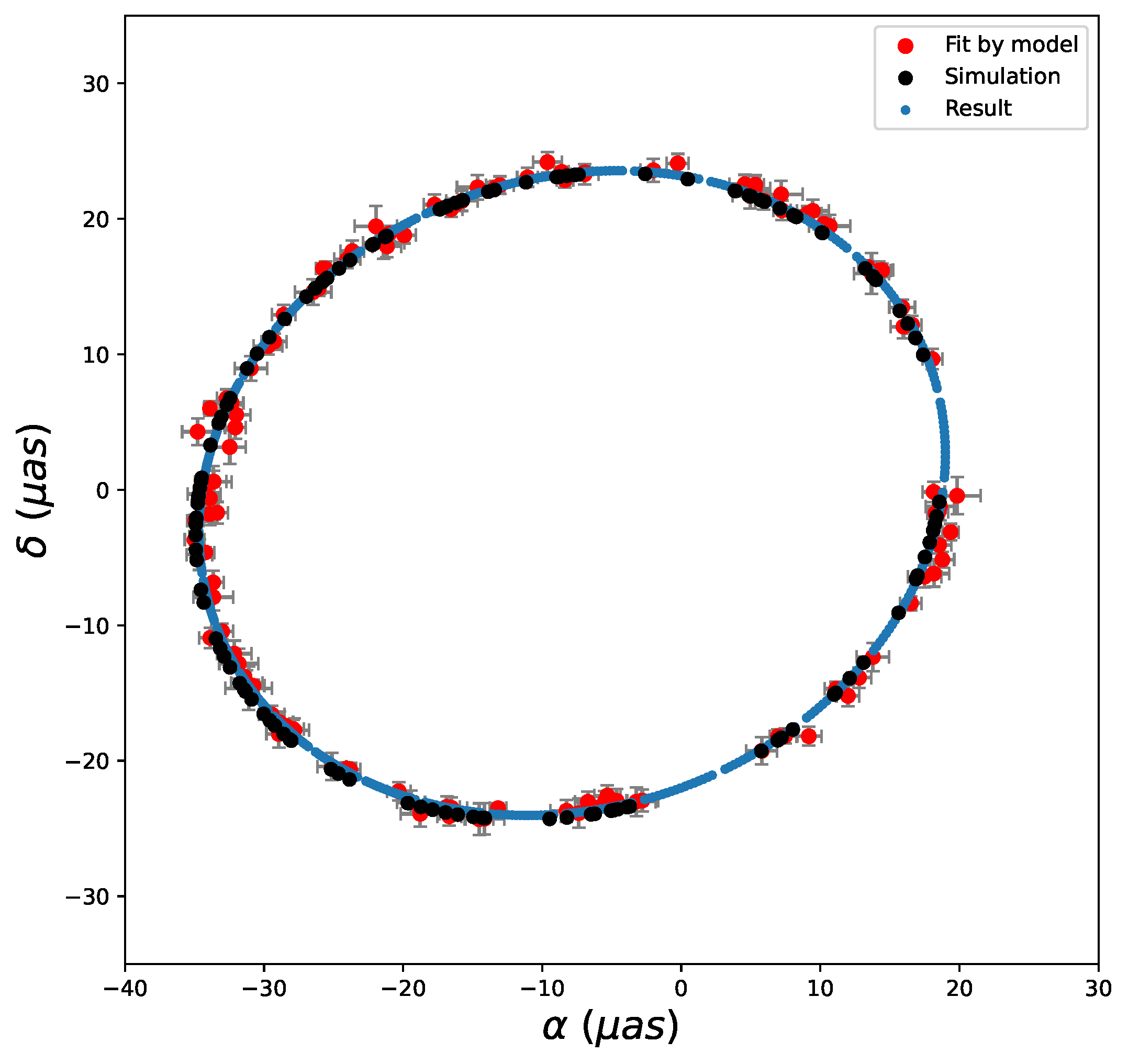}
		\includegraphics[width=0.32\textwidth]{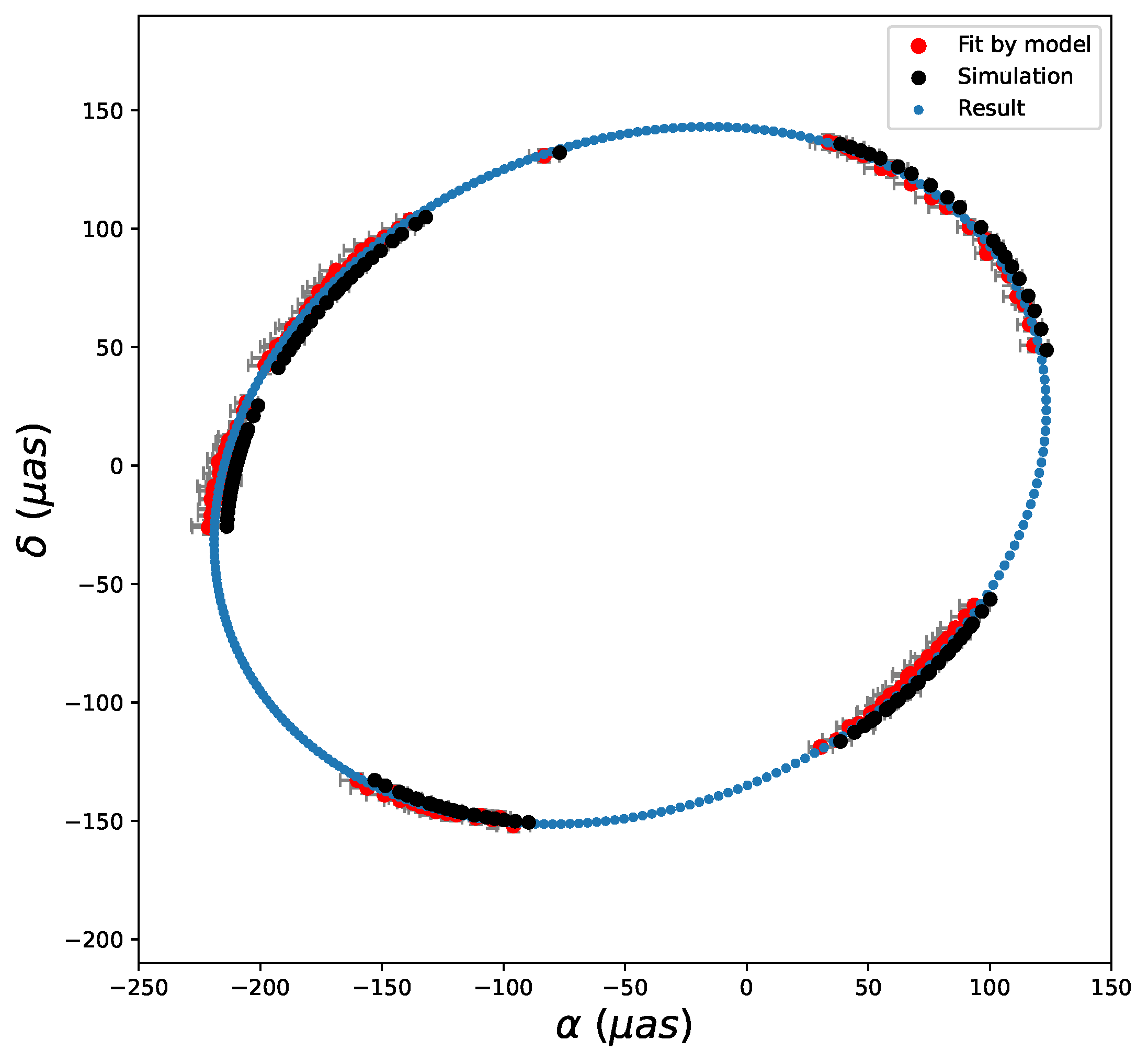}
		\caption{
			Fitting of exoplanets.
			(a) The target star is HD 88230, with a hypothetical Earth-like planet placed at a distance of 1 AU from the star.
			(b) The target star is again HD 88230, with a hypothetical warm Jupiter placed at a distance of 0.1 AU.
			(c) The target star is HD 147513, which is known to host the detected Jovian planet HD 147513 b.
			The angular distance measurement accuracy is $1~\mu \rm as$.
			The black points represent the simulated planetary parameters, while the red points show the residuals after removing proper motion and parallax through fitting.
			The blue points depict the stellar motion orbit induced by planetary perturbations, derived from a Keplerian fit to the residuals.
		}
		\label{Exoplanet}
	\end{figure}

	Within the HD 147513 field of view, nine reference stars brighter than magnitude 12 are observed 153 times over five years. Using the same approach as for Earth-like planets, we simulate the orbital motion and obtain angular separation measurements between the target and reference stars. Results after fitting the model are shown in Figure~\ref{Exoplanet}, and Keplerian fitting yields the planetary parameters listed in Table~\ref{planetfit}.
	
	Assuming an astrometric precision of $1~\mu \rm as$, the signal induced by a Jovian planet is substantially larger than that of an Earth-like planet and can be robustly fitted. Stronger signals, such as from larger semi-major axes, allow the model to recover orbital information more accurately, demonstrating its capability in detecting Jovian planets.
	
	\begin{figure}
		\centering
		\includegraphics[width=0.32\textwidth]{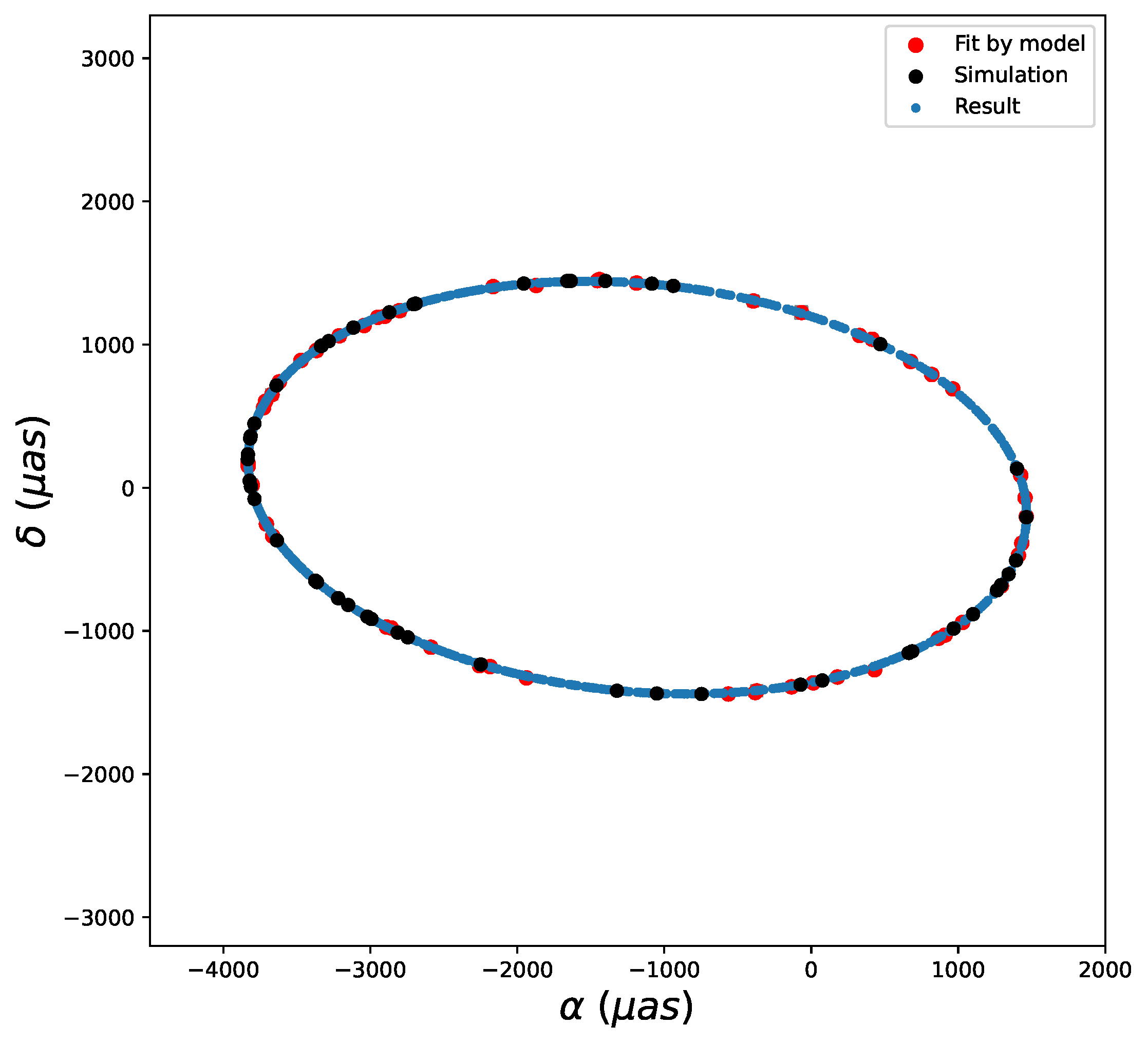}
		\includegraphics[width=0.32\textwidth]{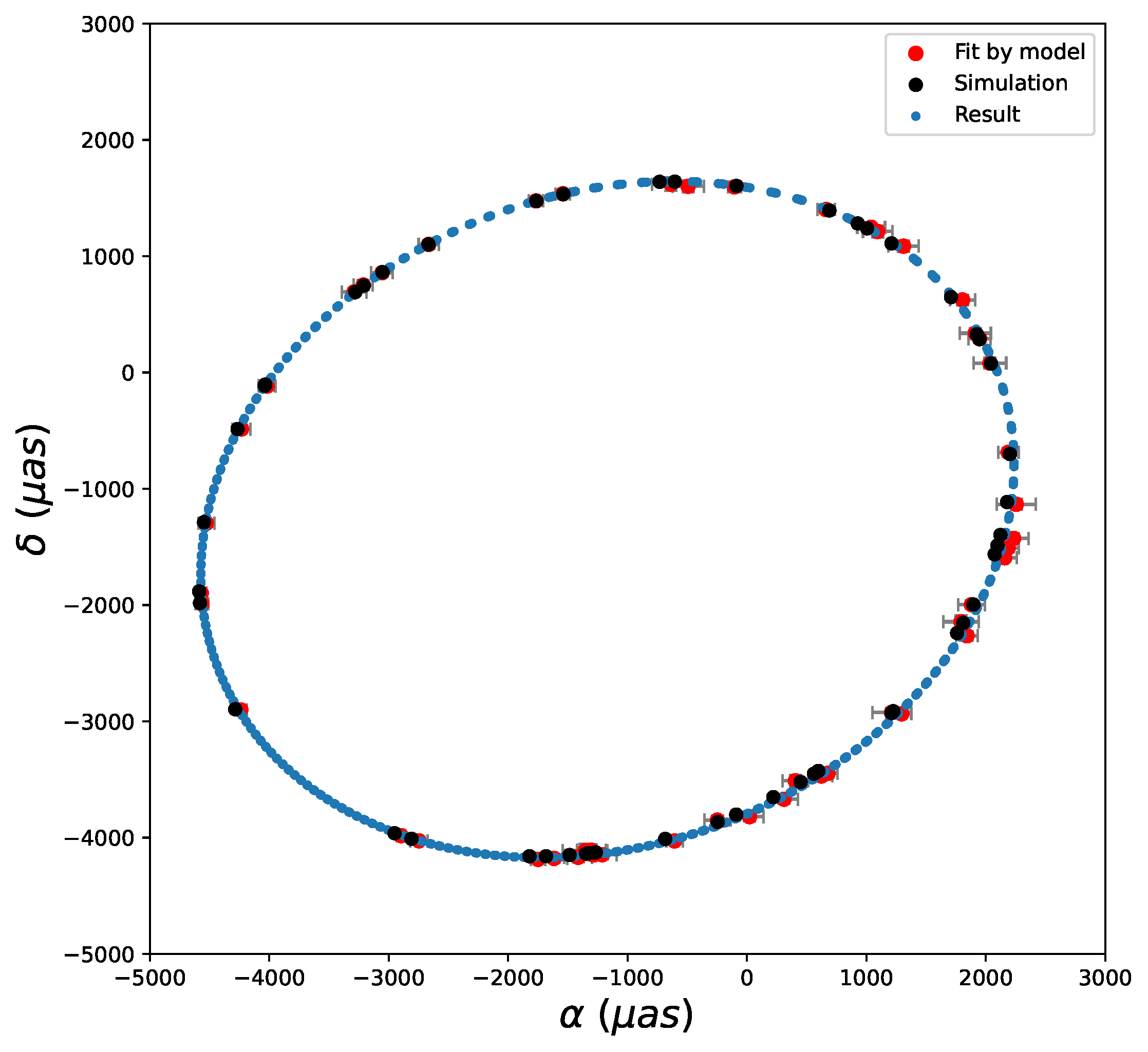}
		\includegraphics[width=0.32\textwidth]{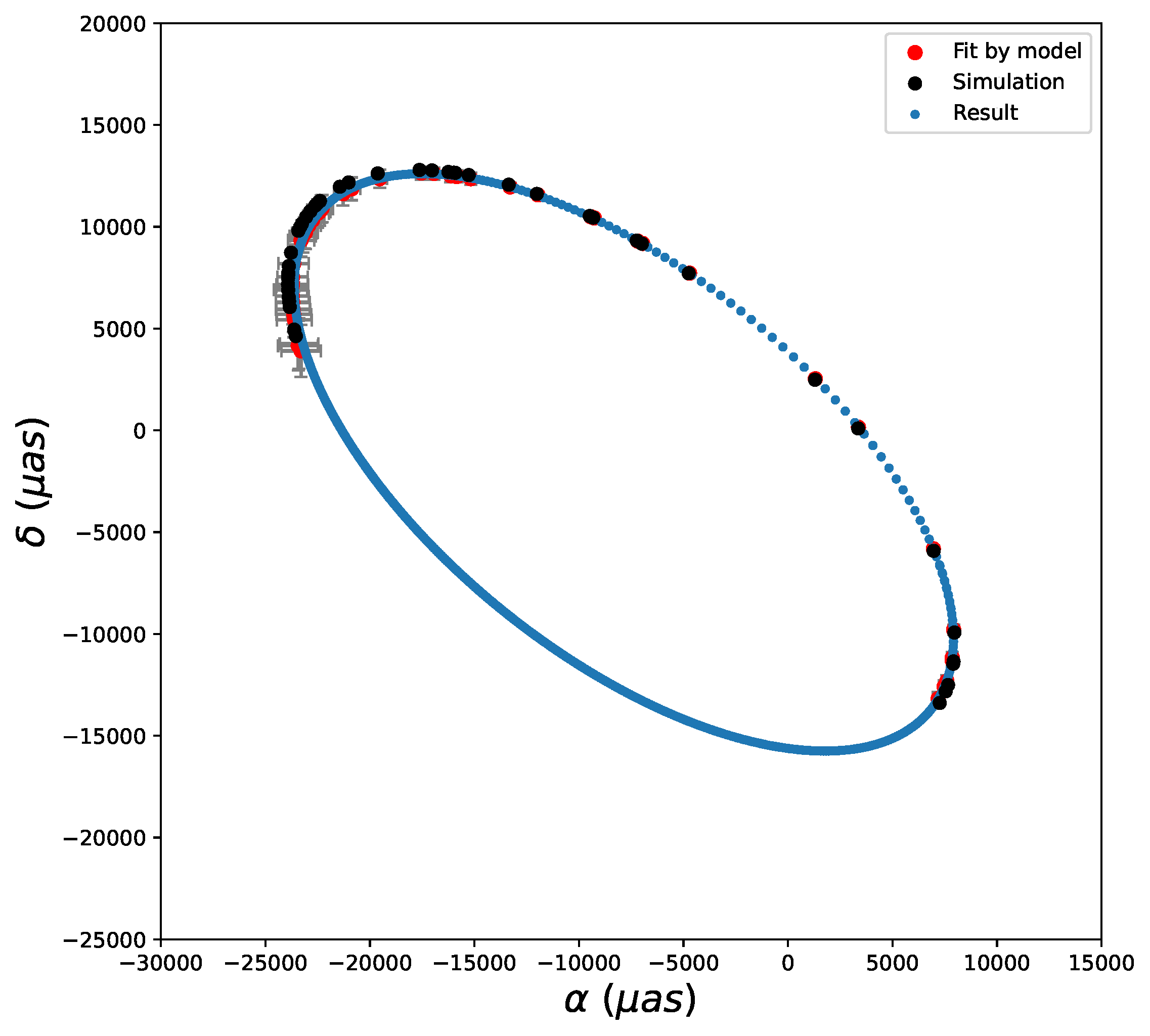}
		\caption{
			Fitting of black holes.
			The three panels correspond to \emph{Gaia} BH1, BH2, and BH3, respectively.
			The angular distance measurement accuracy is $20~\mu \rm as$.
			The black points denote the simulated orbital parameters, while the red points represent the residuals after fitting and removing the proper motion and parallax.
			The blue points represent the stellar motion orbit induced by the gravitational influence of the black hole, obtained through a Keplerian fit to the residuals.
		}
		\label{BHfig}
	\end{figure}
	
	\begin{table*}
		\centering
		\caption{The fitting results for Earth-like planets, Jovian planet and black holes.\label{planetfit}}
		\begin{tabular}{ccccccc}
			\hline
			Parameter & \multicolumn{2}{c}{Earth-like planet} & \multicolumn{2}{c}{Warm Jupiter} & \multicolumn{2}{c}{HD 147513 b$^{a}$} \\
			\cline{2-7}
			& Fitting values & Given values & Fitting values & Given values & Fitting values & Given values \\
			\hline
			$\cos i$ & $0.708^{+0.118}_{-0.077}$ & 0.867 & $0.866^{+0.005}_{-0.006}$ & 0.867 & $0.802^{+0.005}_{-0.005}$ & 0.80 \\
			$e$ & $0.265^{+0.066}_{-0.046}$ & 0.300 & $0.293^{+0.002}_{-0.002}$ & 0.300 & $0.281^{+0.002}_{-0.002}$ & 0.26 \\
			$\Omega$ (deg) & $27.288^{+16.168}_{-11.555}$ & 60.0 & $60.936^{+0.973}_{-1.651}$ & 60.0 & $60.008^{+0.613}_{-0.581}$ & 60.0 \\
			$M_{0}$ (deg) & $80.917^{+13.499}_{-21.376}$ & 45.0 & $46.663^{+0.380}_{-0.406}$ & 45.0 & $228.379^{+0.286}_{-0.275}$ & 228.304 \\
			$\omega$ (deg) & $27.972^{+18.907}_{-12.652}$ & 30.0 & $30.809^{+1.500}_{-1.140}$ & 30.0 & $30.180^{+0.722}_{-0.744}$ & 30.0 \\
			Mass ($M_{\oplus}$) & $0.964^{+0.071}_{-0.046}$ & 1.0 & $299.118^{+1.763}_{-0.615}$ & 300.0 & $629.330^{+2.501}_{-2.710}$ & 625.400 \\
			Period (days) & $425.091^{+9.285}_{-8.148}$ & 447.67 & $14.157^{+0.0001}_{-0.0001}$ & 14.16 & $529.538^{+0.172}_{-0.173}$ & 529.436 \\
			\hline
			Parameter & \multicolumn{2}{c}{BH1$^{b}$} & \multicolumn{2}{c}{BH2$^{c}$} & \multicolumn{2}{c}{BH3$^{d}$} \\
			\cline{2-7}
			& Fitting values & Given values & Fitting values & Given values & Fitting values & Given values \\
			\hline
			$\cos i$ & $-0.595^{+0.001}_{-0.001}$ & -0.596 & $0.818^{+0.005}_{-0.002}$ & 0.820 & $-0.359^{+0.003}_{-0.004}$ & -0.352 \\
			$e$ & $0.451^{+0.001}_{-0.001}$ & 0.451 & $0.515^{+0.002}_{-0.001}$ & 0.518 & $0.722^{+0.005}_{-0.004}$ & 0.729 \\
			$\Omega$ (deg) & $97.782^{+0.113}_{-0.112}$ & 97.81 & $267.722^{+0.430}_{-0.989}$ & 266.9 & $136.103^{+0.240}_{-0.231}$ & 136.24 \\
			$M_{0}$ (deg) & $30.072^{+0.146}_{-0.146}$ & 30.0 & $29.062^{+0.334}_{-0.339}$ & 30.0 & $30.828^{+0.448}_{-0.666}$ & 30.0 \\
			$\omega$ (deg) & $12.859^{+0.204}_{-0.202}$ & 12.8 & $129.291^{+1.099}_{-0.482}$ & 130.9 & $77.087^{+0.561}_{-0.485}$ & 77.34 \\
			Mass ($M_{\odot}$) & $9.694^{+0.023}_{-0.022}$ & 9.62 & $8.485^{+0.073}_{-0.121}$ & 8.94 & $33.566^{+0.346}_{-0.349}$ & 32.70 \\
			Period (days) & $185.594^{+0.011}_{-0.011}$ & 185.59 & $1278.864^{+0.968}_{-0.987}$ & 1276.7 & $4128.321^{+96.837}_{-62.057}$ & 4253.1 \\
			\hline
		\end{tabular}
		
		\vspace{2mm}
		{\small
			References:
			(a) \cite{Mayor2004};
			(b) \cite{El-Badry2023a};
			(c) \cite{El-Badry2023b};
			(d) \cite{GaiaCollaboration2024}.
		}
	\end{table*}
	
	\subsection{Black Hole}
		
	Past black hole detections have primarily relied on indirect electromagnetic evidence, such as X-rays from binary star systems \citep{Remillard2006, Corral-Santana2016}. Astrometric methods provide an alternative, measuring minute orbital shifts of stars to infer the gravitational influence, mass, and orbital properties of black holes \citep{Ghez2008, Gillessen2009}. This approach is especially suitable for isolated black holes and low-mass dark objects. Recently, three possible black holes -- \emph{Gaia} BH1, BH2, and BH3 -- have been identified via their influence on nearby stars \citep{El-Badry2023a, El-Badry2023b, GaiaCollaboration2024}, marking an important step in revealing the distribution of dark objects.
	
	The CHES target catalog does not include potential optical black hole targets, but CHES's high-precision astrometry, including cross-validation among reference stars, may enable identification of such targets.
	
	To test the model's applicability, we selected BH1, BH2, and BH3 optical targets. Their \emph{Gaia} magnitudes are 13.77, 12.28, and 11.23. The eight brightest stars in each field were chosen as references, following the same criteria as for CHES targets. We simulated 50 random observations over five years, adopting \emph{Gaia} DR3 observational errors of $20~\mu \rm as$ \citep{GaiaCollaboration2023}. Processed angular separation data are shown in Figure~\ref{BHfig} and Table~\ref{planetfit}.
	
\begin{figure}
		\centering
		\includegraphics[width=0.45\textwidth, angle=0]{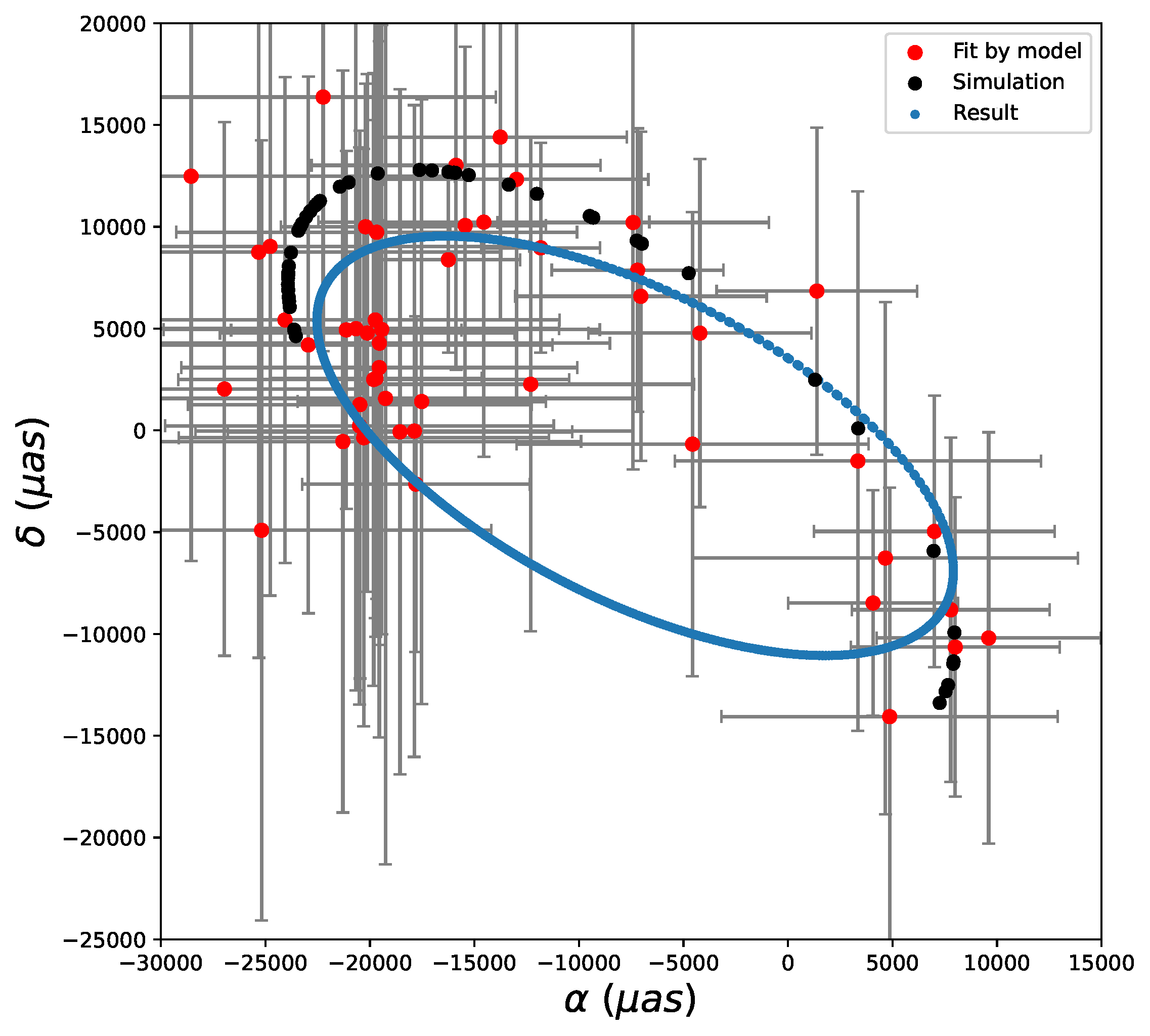}
		\caption{Simulated astrometric observations and orbital fitting for \emph{Gaia} BH3. The astrometric precision of angular separation measurements is referenced from the \textit{Kepler} calibration data, with a relative accuracy of 4~mas. The black points indicate the simulated astrometric data, while the red points represent the residuals after removing the stellar proper motion and parallax. The blue curve shows the stellar orbital motion induced by the gravitational influence of the black hole, derived from a Keplerian fit to the residuals.	
		}
		\label{fig:BH3k}
	\end{figure}

	Additionally, we tested using the photometric satellite \textit{Kepler}, whose angular separation precision is $\sim 4$~mas \citep{Monet2010}. Using this as observational uncertainty, we generated angular separations analogous to the Gaia BH3 simulation, with results in Figure~\ref{fig:BH3k} and Table~\ref{tab:BH3k}.
	
	These simulations demonstrate that the proposed model can be applied to photometric missions. While such missions are not designed for astrometry and derived parameters are less precise than from dedicated missions like \emph{Gaia}, sufficiently high relative angular separation precision can reveal potential astrometric signals. With micro-arcsecond precision, detection capability approaches that of Figure~\ref{BHfig} and Table~\ref{planetfit}.

	\begin{table}
		\centering
		\caption{Fitting results for \emph{Gaia} BH3 using Kepler calibration precision.\label{tab:BH3k}}
		\begin{tabular}{ccc}
			\hline
			Parameter & Fitting values & Given values \\
			\hline
			$\cos i$ & $-0.384^{+0.113}_{-0.079}$ & $-0.352$ \\
			$e$ & $0.657^{+0.101}_{-0.122}$ & $0.729$ \\
			$\Omega$ (deg) & $124.91^{+9.970}_{-8.953}$ & $136.24$ \\
			$M_{0}$ (deg) & $32.598^{+10.976}_{-7.650}$ & $30.0$ \\
			$\omega$ (deg) & $66.062^{+18.338}_{-20.011}$ & $77.34$ \\
			Mass ($M_{\odot}$) & $23.600^{+4.488}_{-6.426}$ & $32.70$ \\
			Period (days) & $3427.559^{+1452.236}_{-761.546}$ & $4253.1$ \\
			\hline
		\end{tabular}
	\end{table}
	
	These examples are provided to demonstrate that our model is also applicable to stellar motion induced by black holes. For black holes that are farther from us, have smaller masses, or are located closer to their optical targets, their signal strength will become weaker. However, with sufficiently high-precision angular separation measurements, this model can enable the detection of black holes. This method can be used to identify perturbation signals arising from black holes, binary stars, or other compact celestial objects, as well as from dynamic effects caused by factors such as dark matter.
	
	\section{Summary and Discussion}
	
	\label{sect:sum}
	
	This work presents a new method for detecting exoplanetary astrometric signals based on variations in angular separation. The proposed approach addresses two major challenges faced by traditional relative astrometry: (1) the difficulty of establishing a stable reference frame due to field rotation at different observing epochs, and (2) the accumulation of catalog errors arising from the intrinsic motion of reference stars and long-term measurement uncertainties. Unlike conventional methods that rely on prior stellar catalogs, our model requires only the one-dimensional temporal variation of angular separations between the target star and multiple reference stars within the field of view. Consequently, the achievable accuracy depends solely on the precision of relative angular separation measurements, without the need to construct a global reference frame.
	
	We have independently modeled the contributions of proper motion, parallax, stellar aberration, gravitational lensing, and planetary perturbations. The complete model achieves sub-microarcsecond accuracy. With microarcsecond-level measurements of angular separations, angular variations along multiple non-collinear directions provided by reference stars enable the reconstruction of the target's two-dimensional orbital motion, allowing the detection of Earth-like planets in the habitable zone. The fitting process also yields astrometric parameters of the target and reference stars, such as total proper motion, radial velocity, and parallax.
	
	In Section~\ref{sec:Num}, we presented several numerical simulations demonstrating that the proposed model can detect habitable-zone Earth-like planets, warm Jupiters, and Jovian planets with microarcsecond precision. Furthermore, the same mathematical framework is applicable to other Keplerian systems, such as binaries and black holes. Verification tests were conducted using simulated \emph{Gaia} BH1, BH2, and BH3 observations.
	
	To further validate the model, we simulated \emph{Gaia} BH3 observations assuming a relative angular separation precision of approximately 4~mas \citep{Monet2010}, consistent with the calibration data of the Kepler mission. The results indicate that gravitational perturbations induced by the black hole remain detectable, confirming that the proposed model can also be applied to photometric missions or other non-astrometric satellites that achieve high-precision relative angular measurements. In the future, as photometric missions reach higher relative astrometric precision, which is generally easier to obtain than absolute precision, it will become feasible to carry out simultaneous photometric and astrometric observations within a local field of view.
	
	Since the observational data consist solely of one-dimensional angular separation variations between stars within the field of view, measurements from different observation programs can be combined within this framework. Angular separations obtained at different times and positions can thus be jointly analyzed to detect stellar motion signals beyond proper motion and parallax.
	
	The current study focuses on systems with a single planet orbiting the target star. Future work will extend this framework to multi-planet systems and to scenarios where reference stars host planetary companions. Moreover, since this method relies on one-dimensional angular variations within the observational plane, we plan to explore its combination with radial-velocity measurements, enabling a more complete three-dimensional reconstruction of stellar motion.
	
	\begin{acknowledgements}
		This work is financially supported by the National Natural Science Foundation of China (grant Nos. 12533011,12033010,12473076), the Strategic Priority Research Program on Space Science of the Chinese Academy of Sciences (Grant No. XDA 15020800), the Foreign Expert Project (grant No. S20240145), and the Foundation of Minor Planets of the Purple Mountain Observatory.		
		
		\textit{Software:}
		\texttt{astropy} \citep{AstropyCollaboration2013},
		\texttt{scipy} \citep{Virtanen2020},
		\texttt{emcee} \citep{Foreman-Mackey2013},
		\texttt{PyMsOfa} \citep{Ji2023},
		\texttt{Nii-C} \citep{Jin2024},
		\texttt{dynesty} \citep{Speagle2020}.
	\end{acknowledgements}

	\clearpage
	\appendix
	\section{Simulation of Parallax Effects}
	\label{A1}
	
	We simulated the positional changes of the target star and reference stars during observations at the L2 point, caused by parallax, and calculated their angular separations. The computational model used is as follows \citep{kovalevsky2011}:
	
	\begin{equation}
		\begin{array}{ll}
			\cos AR =&  \cos (\frac{\pi}{2} - \delta ) \cos(\frac{\pi}{2} - \delta_{\odot })\\
			&+\sin(\frac{\pi}{2}-\delta)\sin(\frac{\pi}{2} - \delta_{\odot })\cos(\alpha -\alpha _{\odot})
		\end{array}
	\end{equation}
	where $AR$ represents the angular separation between the star and the sun. $\delta$ and $\alpha$ are the equatorial coordinates of the star. $\delta_{odot}$ and $\alpha_{\odot}$ are the equatorial coordinates of the sun.
	
	In three dimensions:
	
	\begin{equation}
		\frac{\sin (AR-p)}{d} =  \frac{\sin b}{D} ,
	\end{equation}
	where $d$ represents distance of the target star from the sun, calculated by the parallax of the star. $b$ represents the angular difference between the position of the target star as observed from the observer's location and from the barycenter of the solar system. $D$ is the distance between the observer and the sun.
	
	Formula for Small Angular Displacement:
	
	\begin{equation}
		\begin{array}{ll}
			k &= -p/\sin AR\\
			\alpha'&= \alpha + k\cdot \sec \delta \cos \delta_{\odot}\sin(\alpha-\alpha_{\odot})\\
			\delta'&=\delta + k \cdot [\sin \delta\cos\delta_{\odot}\cos(\alpha -\alpha _{\odot})-\cos\delta \sin\delta_{\odot}],
		\end{array}
	\end{equation}
	where $k$ represent the coefficient of the formula for small angular displacement. $\alpha'$ and $\delta'$ are the equatorial coordinates of the target star after the parallax treatment.
	
	\section{Satellite Orbit Simulation Model}
	\label{app:orbit}
	
	To estimate the aberration and gravitational lensing effects in the simulated observations, we require realistic estimates of the satellite's instantaneous position and velocity with respect to the solar system barycenter. The spacecraft is assumed to operate near the Sun-Earth L$_2$ Lagrange point, where many modern space-based missions such as \textit{Gaia}, \textit{JWST}, and the proposed \textit{CHES} adopt quasi-periodic halo or Lissajous orbits for thermal stability and continuous anti-Sun pointing \citep{deBruijne2012,Gardner2023,Ji2022}.
	
	In the simulation, the Earth follows a circular heliocentric orbit with a radius of 1~AU and a velocity of 29.78~km~s$^{-1}$. The L$_2$ point is located approximately 0.01~AU (1.5$\times$10$^6$~km) beyond Earth in the anti-Sun direction. The satellite motion around L$_2$ is represented by a simplified three-dimensional halo orbit with an amplitude of 0.0017~AU (2.5$\times$10$^5$~km) and a period of 0.5~yr, consistent with typical L$_2$ trajectories of current astrometric and exoplanet missions.
	
	The barycentric position, $\mathbf{r}_\mathrm{sat}(t)$, is obtained by superposing Earth's orbital motion and the periodic oscillations around L$_2$, while the velocity, $\mathbf{v}_\mathrm{sat}(t)$, includes both the co-moving orbital component and the smaller halo-induced variations. This approximate kinematic model captures the essential geometry required for evaluating relativistic effects at the microarcsecond level, without the need for a full $N$-body dynamical integration.
	
	\section{Summary of the Astrometric Model}
	\label{app:model_summary}
	\begin{equation}
		\begin{split}
			E - e \sin E &= \frac{2 \pi}{P} t - M_0,\\[2pt]
			l_{pl} &= D_1 (\cos E - e) + D_2 \sqrt{1-e^2}\, \sin E,\\[2pt]
			\varpi_i &= \varpi_{i0} + V_{ri} \Delta t, \\
			a_{\mu i} &= -2 \frac{\mu_i V_{ri}}{\varpi_i}, \\
			AS_1 &= m_1 + u_1 \Delta t + a_{\mu 1} \Delta t^2
			+ l_{pl}(t),\\
			AS_2 &= m_2 + u_2 \Delta t + a_{\mu 2} \Delta t^2 ,\\[2pt]
			r_i &= \varpi_i
			\sqrt{\frac{\sin^2 \beta_i}{\cos^2 p_i + \sin^2 \beta_i \sin^2 p_i}}, \\
			\cot p_i &= \sin \beta_i \cot q_i,\\[2pt]
			\theta_{i+6} &= \pi - p_i - (w_i + u_{wi} \Delta t),\\
			\cot \theta_{i+8} &= \frac{1}{\sin \theta_{i+6}}
			\Big(\sin AS_i \cot r_i - \cos AS_i \cos \theta_{i+6}\Big),\\[2pt]
			\cos AS'_i &= \cos AS_i \cos r_i + \sin AS_i \sin r_i \cos \theta_{i+6}, \\[2pt]
			\theta_{11} &= \gamma + \text{sign}_1 \theta_9 + \text{sign}_2 \theta_{10},\\[2pt]
			\cos l_{pm+plx} &= \cos AS'_1 \cos AS'_2
			+ \sin AS'_1 \sin AS'_2 \cos \theta_{11},\\
			l_{\rm mod} &\approx k_{ab} k_{gl} l_{pm+plx},\\
			i&=1,2.
		\end{split}
		\label{eq:fullsystem}
	\end{equation}
	
	The above sections present the complete modeling of the variation in the observed angular separation between the target stars. The mathematical formulation of the model is given in Eq.~\ref{eq:fullsystem}.
	
	In this framework, the planetary perturbation acts directly on the apparent position of the star on the celestial sphere, and is therefore treated together with the intrinsic stellar proper motion at the initial stage of the modeling. Based on the a priori known ecliptic coordinates of the Sun, the coefficients $p_i$ and $q_i$ are determined, allowing the parallax contribution to the angular separation to be properly accounted for.
	
	In addition to the orbital parameters associated with the planetary perturbation, the main astrometric parameters to be solved include:
	the angular distance between each star and the fixed point defined by the intersection of their proper-motion directions ($m_i$),
	the total proper-motion amplitude ($u_i$),
	the position angle of the proper-motion direction ($\gamma_i$),
	the radial velocity ($V_{r,i}$),
	and the stellar parallax ($\varpi_i$).
	These parameters collectively describe the kinematic and geometric properties of the system as observed in barycentric coordinates.
	
	Furthermore, the final observed angular separation is affected by stellar aberration and gravitational light deflection, both determined by the instantaneous position and velocity of the observing satellite. The overall accuracy of the model reaches the microarcsecond level, while the precision of the derived parameters ultimately depends on the measurement accuracy of the angular separation, which is determined by the instrument performance and observation quality.
	
	\bibliographystyle{raa}
	\bibliography{ms}
	
	\label{lastpage}
	
\end{document}